\documentclass[twocolumn]{aastex631}

\defcitealias{Yavetz2021}{Paper I}	

\submitjournal{ApJ}

\shorttitle{Separatrix Divergence II}
\shortauthors{Yavetz et al.}

\begin{document}

\title{Stream Fanning and Bifurcations: Observable Signatures of Resonances in Stellar Stream Morphology}

\correspondingauthor{Tomer Yavetz}
\email{tyavetz@ias.edu}

\author[0000-0001-6952-5364]{Tomer D. Yavetz}
\affiliation{Institute for Advanced Study, Einstein Drive, Princeton, NJ 08540, USA}
\affiliation{Department of Astronomy, Columbia University, 550 West 120th Street, New York, NY 10027, USA}

\author[0000-0001-6244-6727]{Kathryn V. Johnston}
\affiliation{Department of Astronomy, Columbia University, 550 West 120th Street, New York, NY 10027, USA}
\affiliation{Center for Computational Astrophysics, Flatiron Institute, 162 5th Avenue, New York City, NY 10010, USA}

\author[0000-0003-0256-5446]{Sarah Pearson}
\altaffiliation{Hubble Fellow}
\affiliation{Center for Cosmology and Particle Physics, Department of Physics, New York University, 726 Broadway, New York, NY 10003, USA}

\author[0000-0003-0872-7098]{Adrian M. Price-Whelan}
\affiliation{Center for Computational Astrophysics, Flatiron Institute, 162 5th Avenue, New York City, NY 10010, USA}

\author[0000-0002-5861-5687]{Chris Hamilton}
\affiliation{Institute for Advanced Study, Einstein Drive, Princeton, NJ 08540, USA}

\begin{abstract}

Recent observations have revealed a trove of unexpected morphological features in many of the Milky Way's stellar streams. Explanations for such features include time-dependent deformations of the Galactic gravitational potential, local disruptions induced by dark matter substructure, and special configurations of the streams' progenitors. In this paper, we study how these morphologies can also arise in certain static, non-spherical gravitational potentials that host a subset of resonantly-trapped orbit families. The transitions, or separatrices, between these orbit families mark abrupt discontinuities in the orbital structure of the potential. We develop a novel numerical approach for measuring the libration frequencies of resonant and near-resonant orbits, and apply it to study the evolution of stellar streams on these orbits. We reveal two distinct morphological features that arise in streams on near-resonant orbits: fans, that come about due to a large spread in the libration frequencies near a separatrix; and bifurcations, that arise when a separatrix splits the orbital distribution of the stellar stream between two (or more) distinct orbit families. We demonstrate that these effects can arise in some Milky Way streams for certain choices of the dark matter halo potential, and discuss how this might be used to probe and constrain the global shape of the Milky Way's gravitational potential.

\end{abstract}


%%%%%%%%%%%%%%%%%%%%%%%%%%%%%%%%%%%%%%%%%%%%%%%%%%

%%%%%%%%%%%%%%%%% BODY OF PAPER %%%%%%%%%%%%%%%%%%

\section{Introduction}

When a globular cluster or dwarf galaxy falls into the potential well of a larger galaxy, tidal forces begin stripping stars from the satellite. The tidally stripped stars are deposited on orbits of the host galaxy that are adjacent, but not identical, to the orbit of the satellite (or the progenitor). The slight differences in orbital characteristics generally cause the stars to spread into long-lived, nearly one-dimensional filaments, known as stellar streams, that approximately trace the orbit of the progenitor \citep{Johnston1996, Helmi1999,Johnston1999,Tremaine1999,Sanders2013}.

Observations of stellar streams have greatly impacted our understanding of the Milky Way (MW) and its dark halo over the past two decades. They have enabled some of the most precise measurements of the Galaxy's mass \citep[e.g.,][]{Law2005,Newberg2010,Gibbons2014,Reino2019} and shape \citep[e.g.,][]{Koposov2010,Law2010,Vera-Ciro2013,Bovy2016}.

Recent ground- and space-based surveys, including SDSS, the Dark Energy Survey (DES), and the \textit{Gaia} Mission, have led to significant advances in the discovery and characterization of the MW's stream population \citep[e.g.,][]{Shipp2018,Mateu2018,Ibata2019,Li2022}. As the catalog of known streams has increased over the past few years, so has the number of morphological features and deviations from the classical picture of tidal stream formation. These features include bifurcations \citep{Belokurov2006,Ramos2021}, gaps and kinks \citep{Price-Whelan2018a,Li2021a}, fans \citep{Sesar2015EVIDENCESTREAM,Price-Whelan2016b,Bonaca2019b,Kuzma2021ForwardTails}, and multiple separate components \citep{Bonaca2019a}.

These features present both an exciting opportunity and a significant challenge for the galactic dynamics community. Streams with complex morphologies may help uncover a much deeper understanding of our Galaxy. Theories and mechanisms to explain these morphologies can now be applied to increasingly robust datasets, with the potential promise of characterizing the precise shape and substructure of our Galaxy. For example, localized gaps or spurs may serve as indicators of interactions with Cold Dark Matter (CDM) substructure \citep{Ibata2002,Johnston2002,Yoon2011,Carlberg2012,Erkal2016a,Bovy2017,Price-Whelan2018a,Bonaca2019}; the abundance and characteristics of the ultra-wide binary population in the stream can be used to probe the smoothness of the potential \citep{Penarrubia2021Creation/destructionStreams}; associations between separate streams through their chemical and dynamical properties can reveal the accretion history of the MW \citep{Johnston1998,Helmi1999a,Koppelman2019,Malhan2022}; and a stream's detailed morphology and orientation can be a strong indicator of the Galaxy's global geometry and its time-dependent nature \citep{Pearson2015,Erkal2016,Bonaca2018,Erkal2019,Vasiliev2021}.

The list of possible reasons for stream morphologies in the previous paragraph demonstrates that the MW is a complex system with multiple evolving components, and each stream is subject to a variety of dynamical effects that all act in concert to create the morphological structure we observe. As a result, disentangling the various possible sources of stream features from each other can often seem like a daunting task, requiring a deep understanding of how each effect acts both in isolation and in parallel with other effects. Identifying specific morphological characteristics that are unique to a certain dynamical mechanism, and developing tests to help differentiate between different types of morphological features in observed streams, will therefore be of high value to the community over the upcoming years and decades.

In what follows, we focus on how halo geometry -- and specifically the existence of resonant orbit families in non-spherical potentials -- affects the morphology of streams. \citet{Pearson2015} identified that the morphology of a single stream can be used to rule out certain configurations of the Galactic potential. \citet{Fardal2015} showed that streams evolving on chaotic orbits in non-spherical potentials can evolve into large, diffuse `fans' instead of the standard thin filaments. Building on these results, \citet{Price-Whelan2016} studied the evolution of streams in triaxial potentials and demonstrated that the density and morphology of streams can be used to trace the dynamical structure of a potential and locate chaotic regions, even when the chaotic timescales\footnote{As measured through either the Maximum Lyapunov Exponent \citep{Lyapunov1992TheMotion,Lichtenberg1992} or the diffusion of fundamental frequencies \citep{Valluri2012}.} associated with the orbits are orders of magnitude greater than the age of the Universe.

In a previous paper \citep[henceforth referred to as Paper I]{Yavetz2021}, we demonstrated how stellar streams can serve as sensitive probes of resonances in a gravitational potential through an effect we labeled \textit{separatrix divergence}. The separatrices, or boundaries between resonant and non-resonant regions, create discontinuities in the orbital structure of the potential, and if a stream's progenitor evolves close enough to a separatrix such that some tidally stripped stars are deposited on the other side of the separatrix, the stream may take on a variety of morphological configurations, including fans and bifurcations. While resonances are related to the existence of chaos, we showed how separatrix divergence can disrupt streams even when none of the individual stars are on chaotic orbits.

In this work, we seek to analyze the morphological manifestations of separatrix divergence, in the context of the aforementioned diversity of morphological features observed in MW streams. We study the unique dynamical characteristics of near-resonant motion and how it relates to observable features in stellar streams. Our overarching goal is to facilitate the characterization of this effect in observational data, and differentiate it from other processes and mechanisms that lead to the disruption of stellar streams.

The structure of this paper is as follows: in Section \ref{sec:tools} we lay out the fundamental setup and tools used throughout the paper to investigate streams evolving near separatrices. Section \ref{sec:res1} focuses on the numerical analysis of near-resonant orbital motion and how it relates to the factors that typically govern the formation of stellar streams. As part of this, we also develop a novel approach based on spectral (frequency) analysis to measure the libration frequencies of near-resonant orbits. We devote Section \ref{sec:res2} to characterizing the morphological features that near-resonant streams exhibit, and explaining their dynamical origin. In Section \ref{sec:obs}, we demonstrate that separatrix divergence can produce observable features in some of the MW's streams within the age of the Galaxy, for certain configurations of the MW's gravitational potential. We discuss the similarities and differences between separatrix divergence and other mechanisms that can lead to the disruption of stellar streams in Section \ref{sec:disc}, and conclude in Section \ref{sec:conc}.

\section{Methods and Tools}
\label{sec:tools}

The primary objective of this paper is to study how the morphology of stellar streams is sensitive to the orbital structure of MW-like potentials. As such, we begin by laying out the three fundamental elements that serve as the basis of this investigation: a realistic representation of an axisymmetric gravitational potential to model the dynamics of the MW (\S\ref{subsec:pot}); the analysis of the orbital structure of this potential through spectral analysis and Surface of Section (SoS) plots (\S\ref{subsec:sos}); and the generation of mock stellar streams (\S\ref{subsec:str}).

\subsection{Choice of Potential}
\label{subsec:pot}

In what follows, we rely on a slightly modified version of the four-component MW potential implemented in the \texttt{gala} package \citep{Price-whelan2017}, based on \texttt{MWPotential2014} \citep{Bovy2014a}. In its original form, this potential consists of a spherical Hernquist nucleus and bulge \citep{Hernquist1990}, a Miyamoto-Nagai disk \citep{Miyamoto1975}, and a spherical NFW halo \citep{Navarro1995}.

In order to investigate a range of potentials in which resonances alter the orbital structure, we replace the spherical NFW halo component with an axisymmetric potential that is aligned with the disk (thus ensuring that the composite potential remains axisymmetric). We follow the \citet{Lee} expression for a triaxial halo potential (with two of the three axes always equal to maintain axisymmetry). For the remainder of this section and in Sections \ref{sec:res1} and \ref{sec:res2} we use an oblate halo potential with an axis ratio of $c/a=0.8$ in density, while in Section \ref{sec:obs}, we vary $c/a$ between 0.6 and 1.4. In all cases, the same rotation curves are maintained in the disk plane. Our choice of axisymmetric potentials also motivates us to use a cylindrical coordinate system throughout this work, parameterized in terms of $(R, \phi, z)$. The full form of the halo's gravitational potential used in this work is presented in Appendix \Ref{sec:app-a}.

\subsection{Orbit Integration and Analysis}
\label{subsec:sos}

Throughout this work, we numerically integrate orbits using an eighth-order Dormand-Prince algorithm \citep{Prince1981}. In order to minimize numerical energy dissipation, we set the integration timestep to at most 1/256 of the orbital period. All orbits studied below conserve energy to a part in $10^{11}$.

Mapping the orbital structure of a potential relies on the classification of orbits into resonant and non-resonant orbit families, a task that can be achieved efficiently through spectral analysis \citep[see, e.g.,][]{Binney1982a}. Spectral analysis has been used to great effect in galactic dynamics for orbit classification and mapping the orbital structure of complex gravitational potentials \citep{Schwarzschild1993, Papaphilippou1996, Papaphilippou1998, Carpintero1998, Valluri1998, Merritt1999}.

Orbits that belong to resonant orbit families are characterized by \textit{commensurable} frequencies (i.e., orbits for which $\displaystyle\sum_{i} n_i\Omega_i = 0$ for a set of small integers $n_i$). In order to differentiate between resonant and non-resonant orbits, we use the Numerical Analysis of Fundamental Frequencies (NAFF) method \citep{Laskar1993,Valluri1998} implemented in the \texttt{SuperFreq} package \citep{Price-Whelan2015} to recover the fundamental frequencies of each orbit. This method involves using Fast Fourier Transforms to identify an orbit's frequencies from complex combinations of the orbital motion along a certain coordinate. Given the axisymmetric nature of the potentials discussed in this work, we find that the most reliable recovery of the relevant orbital frequencies $(\Omega_R, \Omega_\phi, \Omega_z)$ is obtained from complex combinations of the motion in Poincar\'{e} symplectic polar coordinates: $R + iv_R$, $z + iv_z$, and $\sqrt{|2\Theta|}(\cos{\phi} + i\sin{\phi})$, where $\Theta = xv_y - yv_x$ \citep{Papaphilippou1996}.

The effective Hamiltonian of an axisymmetric potential $\Phi(R,z)$ is
\begin{equation}
    \label{eq:eff_ham}
    H_\mathrm{eff} = \frac{1}{2}(p_R^2 + p_z^2) + \Phi_\mathrm{eff}(R,z) \ ,
\end{equation}
where $p_R$ and $p_z$ are the conjugate momenta (we use $p_i$ and $v_i$ interchangeably for studying the motion of test particles), and $\Phi_\mathrm{eff}$ is the effective potential:
\begin{equation}
    \label{eq:eff_pot}
    \Phi_\mathrm{eff}(R,z) \equiv \Phi(R,z) + \frac{L_z^2}{2R^2} \ ,
\end{equation}
with $L_z$ representing the (conserved) angular momentum about the $z$-axis. The relevant phase-space can therefore be reduced to the four-dimensional manifold consisting of $R$, $z$, and their conjugate momenta. Plotting the consequents of the motion in a certain coordinate -- e.g., the values of $R$ and $v_R$ when $z=0$ and $v_z>0$ -- allows us to further reduce the dimensionality of the problem and visualize the orbit in a two-dimensional plane. Orbital representations such as this are known as Surface of Section (SoS) plots or Poincar\'e maps, and we make extensive use of them throughout this work to analyze and visualize the orbital structure of the Galactic potential.

\begin{figure*}
    \includegraphics[width=\textwidth]{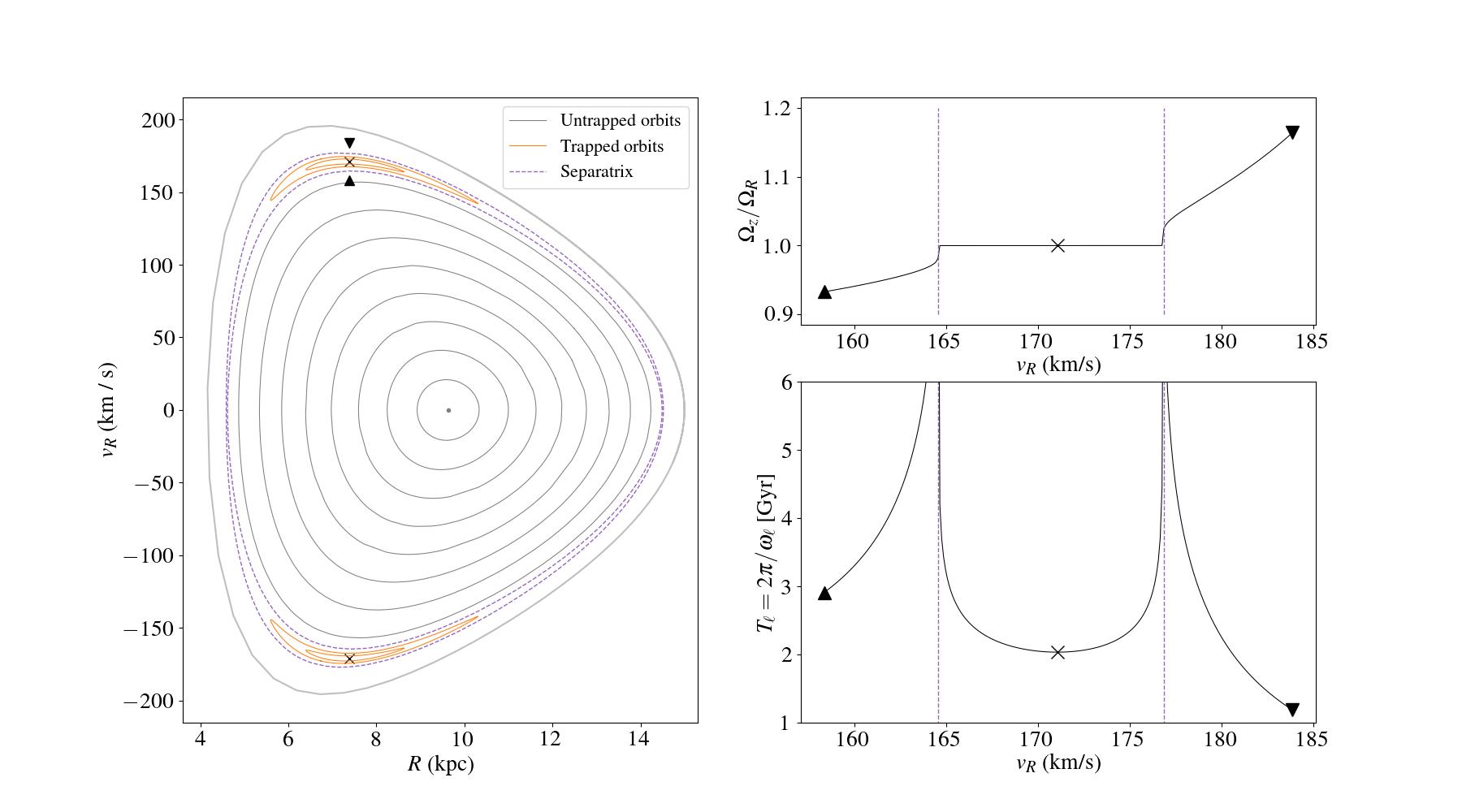}
    \caption{\textit{Left}: a Surface of Section (SoS) plot for the potential described in \S\ref{subsec:pot}, with $E=-(342.5 \mathrm{ km/s})^2$ and $L_z = 1,598 \mathrm{ kpc km/s}$ (in the chosen potential, these values are equivalent to the total energy and 70\% of the total angular momentum of a particle on a circular orbit at $R=10$ kpc, orbiting in the Galactic plane). The SoS is generated by plotting the consequents of 16 orbits, i.e., the values of $R$ and $v_R$ every time $z=0$ and $v_z>0$. The gray contours represent untrapped orbits and the orange contours represent orbits trapped at the $\Omega_z : \Omega_R = 1:1$ resonance. The parent orbits of the resonance are marked by black `$\times$'s. The separatrix, marking the transition between resonant and non-resonant orbits, is plotted in the purple dashed lines. \textit{Upper right}: The frequency ratio of 256 equally-spaced orbits between the two black triangles in the left-hand panel (the initial conditions of the orbits are equally spaced in $v_R$, with fixed values of $E$, $L_z$, $R = 7.4$ kpc, and $z=0$ kpc). The orientation of the two black triangles is preserved across all three panels to assist in identifying the orbits. The flat segment with $\Omega_z / \Omega_R = 1$ corresponds to the resonantly trapped region in the left-hand panel (i.e., the orange contours), and the vertical purple lines mark the separatrix. The frequencies were calculated as described in \S\ref{subsec:sos}, over integration windows of 25 Gyr (long integration times are necessary to smooth over the resonant librations). \textit{Lower right}: The libration period $T_\ell$ ($=2\pi / \omega_\ell$) for each of the orbits in the upper panel, calculated using the SoS data as described in \S\ref{subsec:spec}. Note how the libration period diverges (the libration frequency goes to zero) at the vertical purple lines representing the separatrices. Note also that while we refer to this as a plot of libration periods, only the points between the separatrices belong to librating orbits, whereas those on either side of the separatrix belong to circulating orbits.}
    \label{fig:01}
\end{figure*}

Regular -- i.e., non-chaotic -- orbits in axisymmetric potentials admit a third isolating integral of motion, in addition to $E$ and $L_z$ \citep{Binney2008}. As a result, the consequents of a regular orbit in the two-dimensional SoS plot lie on a one-dimensional invariant curve, as shown in the SoS plot in the left-hand panel of Figure \ref{fig:01}. Critically, each point (and therefore each invariant curve) on the SoS can only belong to a single, distinct orbit at a given energy and angular momentum.

\subsection{Generation of Mock Stellar Streams}
\label{subsec:str}

In order to simulate the behavior of a stellar stream, we begin by integrating the orbit of the stream's progenitor, as described in \S\ref{subsec:sos}. Approximately once per Myr, we ``release'' two test particles with positions and velocities drawn from Gaussian distributions centered on the progenitor's Lagrange points (for a full description of the distribution function used to generate the mock streams, see \citealt{Fardal2015}, or the online documentation of the \texttt{gala} package).

We integrate the orbit of each particle in the chosen MW potential, while also including the perturbation caused by the progenitor, which we model using a Plummer density profile with scale lengths ranging from 4 to 100 parsecs (in practice this small perturbation has no effect on the outcome, and the same results are achieved when the progenitor's potential is ignored). For each simulated stream, we end up with 12,000 particles by the 6 Gyr mark. With this ensemble of orbits in hand, we can apply the tools from \S\ref{subsec:sos} to classify the nature of each particle's orbit.

\section{Results I: Exploring Orbital Properties Near Resonances}
\label{sec:res1}

The mechanism of separatrix divergence, first introduced in \citetalias{Yavetz2021}, describes the rapid diffusion of two or more particles orbiting close to the boundaries between resonant and non-resonant orbit families. Accurate characterization of orbital dynamics near resonances is therefore central to the understanding of separatrix divergence. We are primarily interested in analyzing the locations of resonances in axisymmetric potentials, what volume of orbital space they occupy, what timescales the librational motion encompasses, and the extent to which they create discontinuities in the orbital structure of the potential, with the ultimate goal of tying these to observable quantities so that the potential itself can be characterized. The classical approach in non-linear dynamics to answering these types of questions invokes secular perturbation theory to approximate the near-resonant motion \citep{Chirikov, Arnold1989, Lichtenberg1992, Mcgill1990, Binney1993, Kaasalainen1994b}. The formalism is reviewed briefly in Appendix \ref{sec:app-b}, but at its heart, it approximates the near resonant motion using the equation of motion for an anharmonic oscillator. This can subsequently be used to evaluate the extent of the resonantly trapped region in action space $I_\mathrm{h}$, as well as the frequency of libration $\omega_\ell$ (following the notation from Appendix \ref{sec:app-b}). However, this approach does a poor job of capturing the orbital behavior close to the separatrix, which happens to be the regime of interest in this work.

In what follows, we develop and apply a non-perturbative, numerical approach based on spectral analysis of orbital fundamental frequencies. We begin with a brief review of the characteristics of near-resonant motion through analysis of SoS plots (\S\ref{subsec:qual}). In \S\ref{subsec:spec} we describe our spectral analysis approach and discuss its advantages and shortcomings compared to the traditional perturbative approach. We conclude this section by comparing the properties of near-resonant motion to the relevant quantities and timescales that govern the formation and evolution of stellar streams (\S\ref{subsec:streams}). 

\subsection{Characteristics of Near-Resonant Orbits}
\label{subsec:qual}

Most non-spherical potentials host a multitude of minor orbit families surrounding a small subset of special orbits with commensurable frequencies, often referred to as the \textit{parent orbits} of these orbit families. In the axisymmetric potentials discussed in this paper, minor orbit families form around orbits whose radial and vertical frequencies are commensurable ($\Omega_z : \Omega_R = m : n$ where $m$ and $n$ are small integers). Oscillations in either $R$ or $z$ in axisymmetric potentials involve changes in the magnitude of the gravitational acceleration, leading to a resonant effect when these oscillations are coupled.\footnote{Note that this is not true for the azimuthal oscillations in an axisymmetric potential, since neither the magnitude of the acceleration nor the effective Hamiltonian (Equation \ref{eq:eff_ham}) have any $\phi$-dependence. Thus, only $\Omega_z : \Omega_R$ commensurabilities lead to resonant trapping, and we need not worry about commensurabilities between the azimuthal frequency $\Omega_\phi$ and either of the other two frequencies.}

In many cases, the resonance will be strong enough to trap a subset of the orbits in the vicinity of the closed orbit and cause them to librate around the parent orbit at a certain libration frequency $\omega_\ell$, thus creating an orbit family with unique orbital characteristics. Notably, the fundamental frequencies of these trapped orbits will \textit{appear} to also be commensurable, if calculated for orbits that were integrated over long periods of time compared to the libration period $T_\ell \equiv 2\pi / \omega_\ell$ (determining the frequencies from shorter segments of these orbits will return frequencies that appear to oscillate around the commensurability).

\begin{figure}
    \includegraphics[width=\columnwidth]{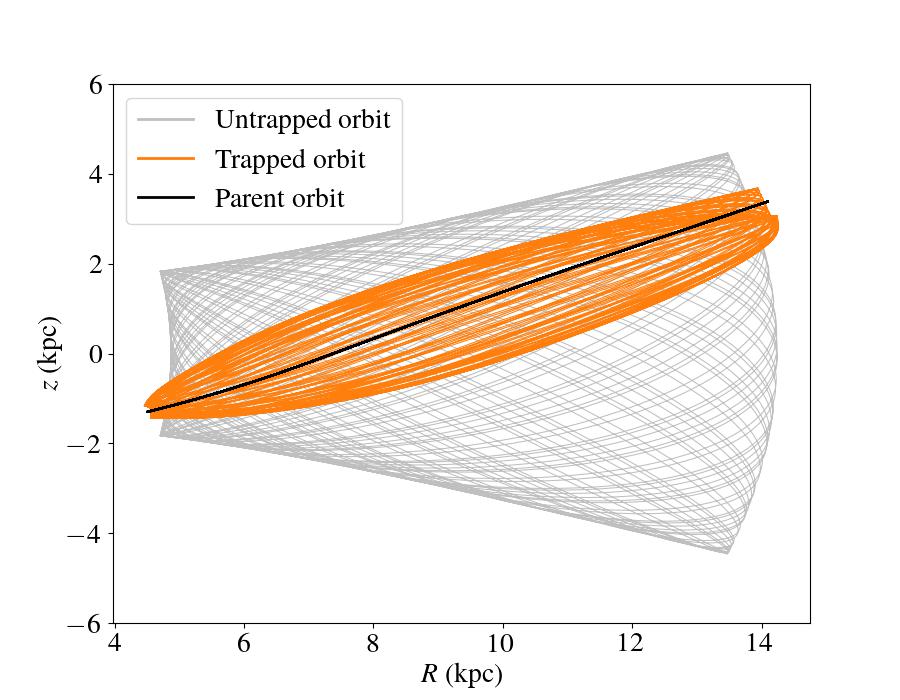}
    \caption{$R-z$ plane projections of three orbits from the SOS shown in Figure \ref{fig:01}. The gray (untrapped) orbit corresponds to the gray contour just inside the smaller branch of the separatrix, and the orange (trapped) orbit corresponds to the larger of the two orange contours in the upper resonant island. The black orbit corresponds to the 1:1 commensurable orbit that parents the trapped orbit family at the 1:1 resonance, represented by the upper black '$\times$' in Figure \ref{fig:01}. No member of the trapped orbit family parented by this commensurable orbit will fill the top-left or the bottom-right areas of the plot like the untrapped orbit does.}
    \label{fig:02}
\end{figure}

The minor orbit family associated with the $\Omega_z : \Omega_R = 1 : 1$ resonance, often referred to as the banana or saucer orbit family, can be seen at the top and bottom of the SoS plot in Figure \ref{fig:01}. While most of the orbits in this SoS plot form concentric closed invariant curves around the gray dot in the middle that represents the shell orbit at $R\approx10$ kpc and $v_R = 0$ km/s, a few of the invariant curves (those plotted in orange) surround the black $\times$ symbol, which represents the commensurable parent orbit of the $1:1$ resonant orbit family. Figure \ref{fig:02} shows the $R-z$ trajectory of an untrapped (gray) and a trapped (orange) orbit from the SoS in Figure \ref{fig:01}, as well as the commensurable parent orbit (in black).\footnote{Note that the two resonant islands at the top and bottom of the plot consist of distinct groups of phase-space orbits; the orbits in the bottom island trace upside-down versions of the bananas traced by the orbits in the top island (equivalent to reflecting the orange orbit in Figure \ref{fig:02} across the $z=0$ line).}

The transition, or $separatrix$, between trapped and untrapped orbits (shown in the purple dashed lines in the panels of Figure \ref{fig:01}) marks an abrupt discontinuity in the orbital structure of the potential; orbits on either side of the separatrix with nearly identical initial conditions belong to two different orbit families and thus have very different orbital characteristics. In more complex potentials, including most triaxial potentials, a chaotic band of orbits surrounds the separatrices. However, we limit our discussion in this work to mildly flattened axisymmetric potentials in which the number of chaotic orbits is negligible, and all orbits studied in this work are regular.\footnote{Axisymmetric potentials can, in some cases, host a few chaotic orbits surrounding the separatrices, but these typically represent a miniscule fraction of orbits in the potential \citep{Pascale}.}

\subsection{Spectral Analysis of Near Resonant Dynamics}
\label{subsec:spec}

To leading order, a resonant orbit family can be well described through two quantities: its extent, or `size', and its libration timescale. In the traditional action-angle formalism, these two quantities can be obtained through secular perturbation theory ($I_\mathrm{h}$ and $\omega_\ell$ as defined in Appendix \ref{sec:app-b}). The perturbation theory approach accurately models the trapped orbits close to the parent orbit of the resonant orbit family, but as one nears the separatrix, the results of perturbation theory begin deviating from reality, and require additional corrections to be made \citep{Kaasalainen1994}. As our main interest lies in studying streams in proximity to separatrices, we develop a complementary, non-perturbative approach, to study these two quantities of interest.

While the location and `size` of a resonant orbit family is readily apparent in SoS plots, its orbital characteristics can be evaluated in a more quantitative fashion through an evaluation of the frequencies of the motion near the resonance. Fourier decomposition of each orbit shown in the SoS in Figure \ref{fig:01} reveals a mostly smooth progression from the central orbit in the plot, where $\Omega_R > \Omega_z$, to the outer curve (the zero-velocity curve), where $\Omega_z > \Omega_R$. However, as shown in the top right panel of Figure \ref{fig:01}, a stark discontinuity appears at $\Omega_z = \Omega_R$, corresponding to the resonantly trapped orbit family (though they are not shown in Figure \ref{fig:01}, similar, smaller discontinuities also appear around higher order resonances). The boundaries of this transition correspond precisely to the separatrix, outside which the frequency ratio jumps back to a value $\neq1$. One may think of this panel as a simplified, one-dimensional version of frequency-frequency plots used to map the orbital structure of a potential \citep[see, e.g.,][p. 260, Figure 3.45]{Binney2008}. Critically, these figures serve as a direct representation of the `size' of the resonantly trapped region in phase-space.

The other quantity of interest related to the near-resonant motion is the libration frequency $\omega_\ell$, which quantifies the rate of a near-resonant orbit's oscillations around the parent orbit. This frequency should be calculable directly from the time-series data of the orbit, in the same manner as any of the other fundamental frequencies.\footnote{It is worth pointing out that near-resonant 3D orbits can still be thought of as having only three independent fundamental frequencies, with $\omega_\ell$ replacing one of the commensurable frequencies (which itself is no longer a linearly independent frequency given the commensurability with another of the fundamental frequencies).} However, $\omega_\ell$ is typically considerably smaller than the other frequencies, so a n{\"a}ive application of NAFF will, in the best case, lead to the recovery of a frequency that is some linear combination of $\omega_\ell$ and one of the other frequencies, or possibly to a failure to detect $\omega_\ell$ at all.

To overcome this, we show that the libration frequency can be reliably recovered from a Fourier decomposition of the SoS data. In other words, rather than using the complex representation of the full integrated orbit, we take the Fourier transform of only the consequents used to plot the SoS. This isolates the librational motion of the near-resonant orbit and allows the NAFF algorithm to efficiently extract the value of $\omega_\ell$. In the bottom right panel of Figure \ref{fig:01}, we plot the value of $T_\ell = 2\pi / \omega_\ell$ corresponding to each orbit evaluated in the top right panel of Figure \ref{fig:01}. 

As expected, the plot exhibits the typical pendulum-like behavior on which the traditional perturbation theory approach is based. Near the center of the resonance, $T_\ell\sim 2$ Gyr, and as one nears the separatrices its value grows rapidly, similar to the period of oscillation of a rigid pendulum with nearly enough energy to reach a vertical configuration (i.e., the unstable balance point at the apex of the allowed circular trajectory).

Note, however, that the values of $T_\ell$ near the separatrices deviate considerably from the motion of a simple anharmonic oscillator. In particular, the behavior on either side of a given separatrix is highly asymmetric in velocity-space; the gradient of the libration time is nearly flat throughout most of the resonantly trapped region ($165$ km/s $< v_R < 177$ km/s), and then rises very rapidly only very near the separatrix, while on the untrapped sides of the separatrix ($v_R < 165$ km/s or $v_r > 177$ km/s), the timescale shrinks much more gradually with distance from the separatrix. As a result, relying on only the first coefficient of the Fourier expansion, as is the norm in secular perturbation theory, can lead to significant inaccuracies near the separatrix. Various higher-order corrections have been proposed to overcome the issues arising from this asymmetry \citep[see, e.g., the modified pendulum equations in][]{Kaasalainen1994}. Our proposed approach using spectral analysis to study the near-resonant motion treats each orbit individually and is only limited by the integration time used to create the input data. It is thus more reliable for recovering the libration frequencies, particularly those near the separatrix.

Unlike secular perturbation theory, our approach only recovers a subset of the information necessary to fully model a resonantly trapped orbit. We do not, for example, recover the extent of trapping in action space, nor the evolution of the slow angle associated with the librating dynamics. However, we have shown how the extent of the resonantly trapped region is well-defined in the SoS plot and through spectral analysis, in a way that can be easily translated to phase-space information. In the next subsection, we will discuss how these findings relate to the properties of stellar streams, as a preliminary step to predicting how separatrix divergence will affect their morphology.

\subsection{Comparing Stellar Streams to Resonances}
\label{subsec:streams}

\begin{figure*}
    \includegraphics[width=\textwidth]{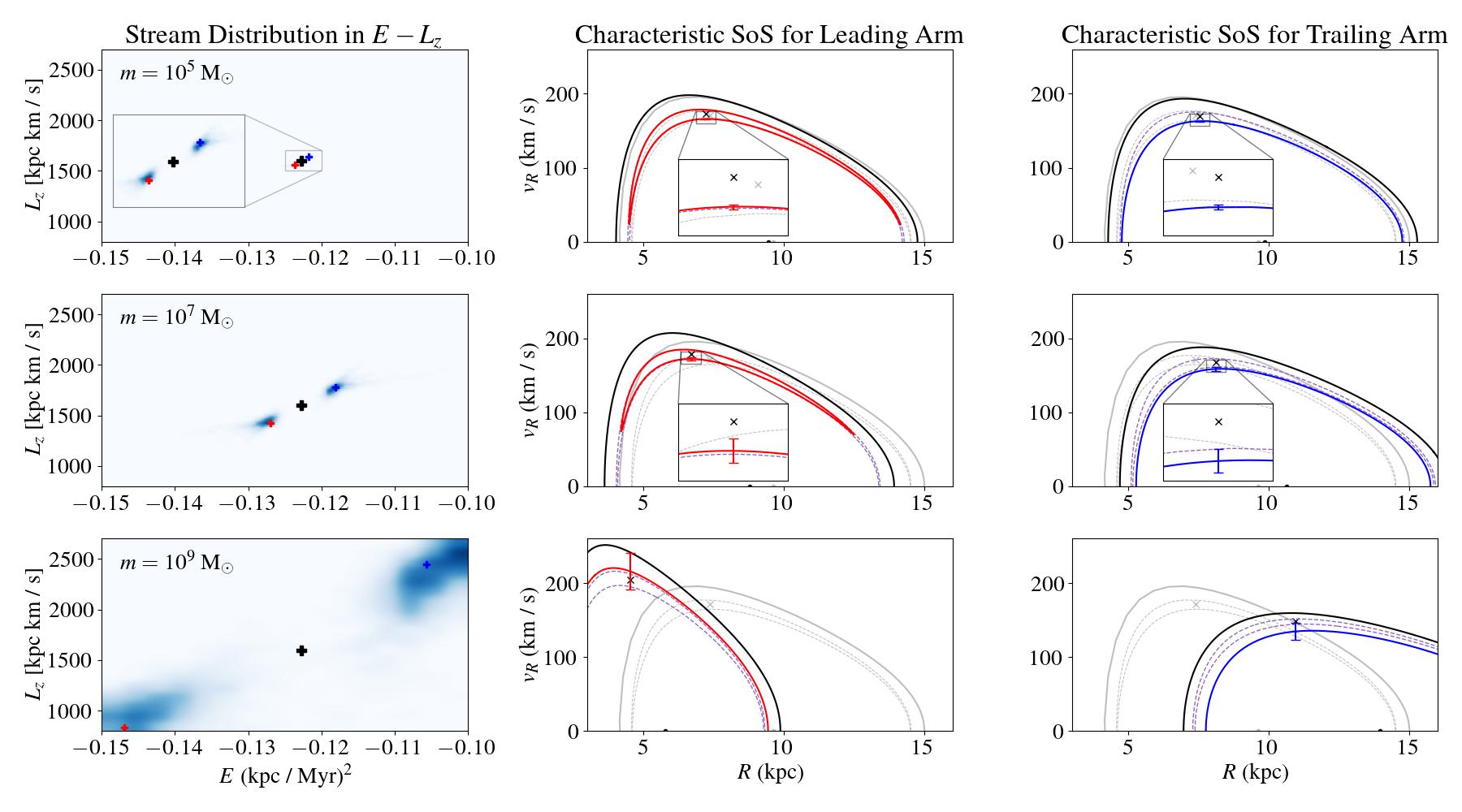}
    \caption{\textit{Left column}: 2D histograms depicting the distribution of stellar streams in $E-L_z$ space for progenitor masses of $10^5$ M$_\odot$ (top row), $10^7$ M$_\odot$ (middle row), and $10^9$ M$_\odot$ (bottom row). The black `+'s in the center mark the energy and angular momentum of the progenitor orbit (which is the same in all three rows), and the two clusters in each figure represent the leading and trailing arms. Each stream is generated as described in \S\ref{subsec:str}, and initialized close to the separatrix. A representative bin in $(E, L_z)$ from the leading (trailing) arm is chosen and marked with a red (blue) `+'. \textit{Middle column}: the SoS for the chosen representative bin from the leading arm, marked by the red '+' in the left-hand panel. The location of the outer SoS orbit, the resonant parent orbit and the separatrix are plotted using the same line styles, markers, and colors as Figure \ref{fig:01}, and the equivalent orbits from the progenitor's SoS (i.e., the exact contours from Figure \ref{fig:01}) are shown in light gray for reference. The median orbit in the selected $(E, L_z)$ bin is plotted with a solid red line, and its spread in the SoS is shown with the error bars, which correspond to $\delta v_\mathrm{sos}$ as defined in \S\ref{subsec:streams}. \textit{Right column}: same as the middle column for the trailing arm, with the median orbit and error bars plotted in blue.}
    \label{fig:03}
\end{figure*}

The aforementioned characteristics of a resonance (`size' and libration time) only supply one half of the picture necessary for understanding separatrix divergence. The other half involves finding a way to compare these quantities to the phase-space volume and phase-mixing time of a stellar stream. In terms of timescales, the regime of interest involves streams whose phase-mixing time $T_\mathrm{pm}$ is comparable to or longer than the libration time $\omega_\ell = 2\pi / T_\ell$, as described in greater detail in \citetalias{Yavetz2021}. To address the phase-space volume comparison, we already demonstrated how SoS plots serve as an excellent tool for visualizing and quantifying the extent of a resonantly trapped region in phase-space. We therefore devote this section to finding a way to describe the size and orbital distribution of stellar streams in SoS plots.

One may gain a first-order intuition for the characteristic size of a stream in an SoS plots by considering the tidal scale of the progenitor with respect to the host halo. In general, a stellar stream can be approximated as a cluster of particles in 6D phase-space that follows a normal distribution with:
\begin{eqnarray}
    \label{eq:sigma_x}
    & \sigma_x \sim \frac{r_p}{\sqrt{3}}\bigg(\frac{m}{M(<r_p)}\bigg)^{\frac{1}{3}} \\ 
    \nonumber & \sim \ 27\ \mathrm{pc}\ \bigg(\frac{r_p}{10\ \mathrm{kpc}}\bigg) \bigg(\frac{m}{10^5\ \mathrm{M}_\odot}\bigg)^{\frac{1}{3}} \bigg(\frac{M(<r_p)}{10^{12}\ \mathrm{M}_\odot}\bigg)^{-\frac{1}{3}} \ ,
\end{eqnarray}
and
\begin{eqnarray}
    \label{eq:sigma_v}
    & \sigma_v \sim \frac{v_p}{\sqrt{3}}\bigg(\frac{m}{M(<r_p)}\bigg)^{\frac{1}{3}} \\ 
    \nonumber & \sim \ 0.54\ \mathrm{km/s}\ \bigg(\frac{v_p}{200\ \mathrm{km/s}}\bigg) \bigg(\frac{m}{10^5\ \mathrm{M}_\odot}\bigg)^{\frac{1}{3}} \bigg(\frac{M(<r_p)}{10^{12}\ \mathrm{M}_\odot}\bigg)^{-\frac{1}{3}} \ ,
\end{eqnarray}
where $\sigma_x$ and $\sigma_v$ represent the spread in position and velocity, $r_p$ and $v_p$ represent the magnitude of the progenitor's position and velocity, $m$ is the mass of the progenitor, and $M(<r_p)$ is the mass of the host halo enclosed within the progenitor's orbit. However, this simple picture fails to take into account some important nuances of stream structure, so in what follows we adopt a slightly more careful approach for assessing the size of a stream in an SoS plot.

A central challenge in visualizing a stream with an SoS plot is that the latter is designed to compare orbits with the same energy and angular momentum. Plotting two orbits with different values of $E$ and/or $L_z$ in the same SoS, can lead to invariant curves that intersect each other, thus violating one of the key concepts of SoS plots. The orbits of stellar stream members cover a range of values in angular momentum and energy, meaning that plotting them together in the same SoS plot can sometimes lead to visually incoherent results.

It is still possible to learn much about the spread of a stream from SoS plots by isolating orbits from the stream in bins of $(E, L_z)$. In the left-hand column of Figure \ref{fig:03}, we show $(E, L_z)$ histograms of three streams with progenitor masses of $10^5$, $10^7$, and $10^9$ M$_\odot$. The approximate mass enclosed within the orbits is $10^{11}$ M$_\odot$, so these progenitors correspond to tidal scales of $m/M\sim10^{-6}$, $10^{-4}$, and 0.01, respectively. The streams are generated as described in \S\ref{subsec:str}, on identical progenitor orbits initiated from a point in phase-space on the lower branch of the separatrix in Figure \ref{fig:01}, with $(R, v_R) \approx (7.4$ kpc$, 164.5$ km/s$)$. Here we have chosen galactic units to give the reader a physical sense of the scales involved, but this is inherently a scale-free problem, where the main quantity of interest is the mass ratio between the progenitor and the host (see Equations \ref{eq:sigma_x} and \ref{eq:sigma_v}, and Appendix \ref{sec:app-b} for a more general argument on the scale-free nature of the problem).

The location of the progenitor orbit is marked in each of the three panels with a black `+', with a large cluster of stars on either side representing the leading and trailing arms. The extent of the stream in $E-L_z$ space grows, as expected, with the mass of the stream (note that we use the same number of test particles to simulate each stream -- we are interested in relative over- or under-densities, or other such features caused by separatrix divergence, so the total number of particles per stream is immaterial, provided it is sufficiently large to produce a good statistical sampling of the stream).

For each stream, we select a representative bin from both the leading and trailing arm, marked with red and blue `+'s in the plot, respectively (the bin size is set to $\sim 10\%$ of the standard deviation of the spread in energy and angular momentum). Since the stars in these bins have approximately equal values of $E$ and $L_z$, we can proceed with plotting them in one SoS plot with fixed $E$ and $L_z$ in order to evaluate the spread of the stream with respect to the extent of a resonantly trapped region. The new SoS's are shown in the middle and right columns for the leading and trailing arms, respectively, and the SoS of the progenitor (the one plotted in Figure \ref{fig:01}) is shown for reference in each of these plots in light gray. The orbit of the median test particle in each bin is plotted in heavy line (red for the leading particle and blue for the trailing particle), to demonstrate how the bin's location in $E-L_z$ space has shifted with respect to the separatrix. In all cases, the leading arm moves `above' the separatrix (i.e., to an orbit with a greater radial action), becoming a resonantly trapped orbit for the $10^5$ and $10^7$ M$_\odot$ cases, and an untrapped orbit higher above the resonantly trapped region in the $10^9$ M$_\odot$ case. Similarly, the trailing arm always ends up centered around an untrapped orbit `below' the separatrix.

Having isolated stream orbits at representative values of $E$ and $L_z$, the final -- and most important -- task is to show the spread of these orbits with respect to the resonance. To achieve this, we analyze the invariant curves of the orbits in each of the selected bins, and record their $v_R$ value when $R$ is identical to the $R$ coordinate of the parent orbit of the resonance (marked with a black `$\times$' in the SoS plots). To compare the size of the stream to the width of the resonantly trapped region, we plot error bars around the median orbit representing the 1-$\sigma$ deviation of this set of $v_R$ values (henceforth, we refer to these spreads as $\delta v_\mathrm{sos}$). In general, we find that $\delta v_\mathrm{sos}$ is smaller than $\sigma_v$ as calculated from Equation \ref{eq:sigma_v} by a factor of $\sim$ 2-3 for each of the three examples shown in Figure \ref{fig:03}. The value of $\delta v_\mathrm{sos}$ remains unchanged if the streams are initialized on nearby resonant or non-resonant orbits instead of the separatrix orbit.

Figure \ref{fig:03}, then, helps to paint a detailed picture of the location and the extent of a stream in comparison to a resonance. In general, we expect a stellar stream to deviate from its typical morphology due to the orbital discontinuity at a separatrix when two conditions are satisfied: (1) the phase-space volume occupied by the stream, as described above, is intersected by a separatrix; (2) the libration time of that resonance is shorter than or comparable to the phase-mixing time of the stream. Studying the morphological properties of streams that satisfy these two conditions is the focus of the next section.

\section{Results II: Stream Morphologies near Separatrices}
\label{sec:res2}

We now study the morphological effects of separatrix divergence on simulated streams. Our goal is to classify the different types of morphological effects that emerge, describe the precise mechanism behind these effects, and understand the relations between streams and resonances (in terms of relative position and size) that produce observable effects. While the focus of this section is characterizing the spatial morphology of streams near resonances, the results extend to other projections of the stream in phase-space, including the stream's velocity distribution.

\begin{figure*}
    \includegraphics[width=\textwidth]{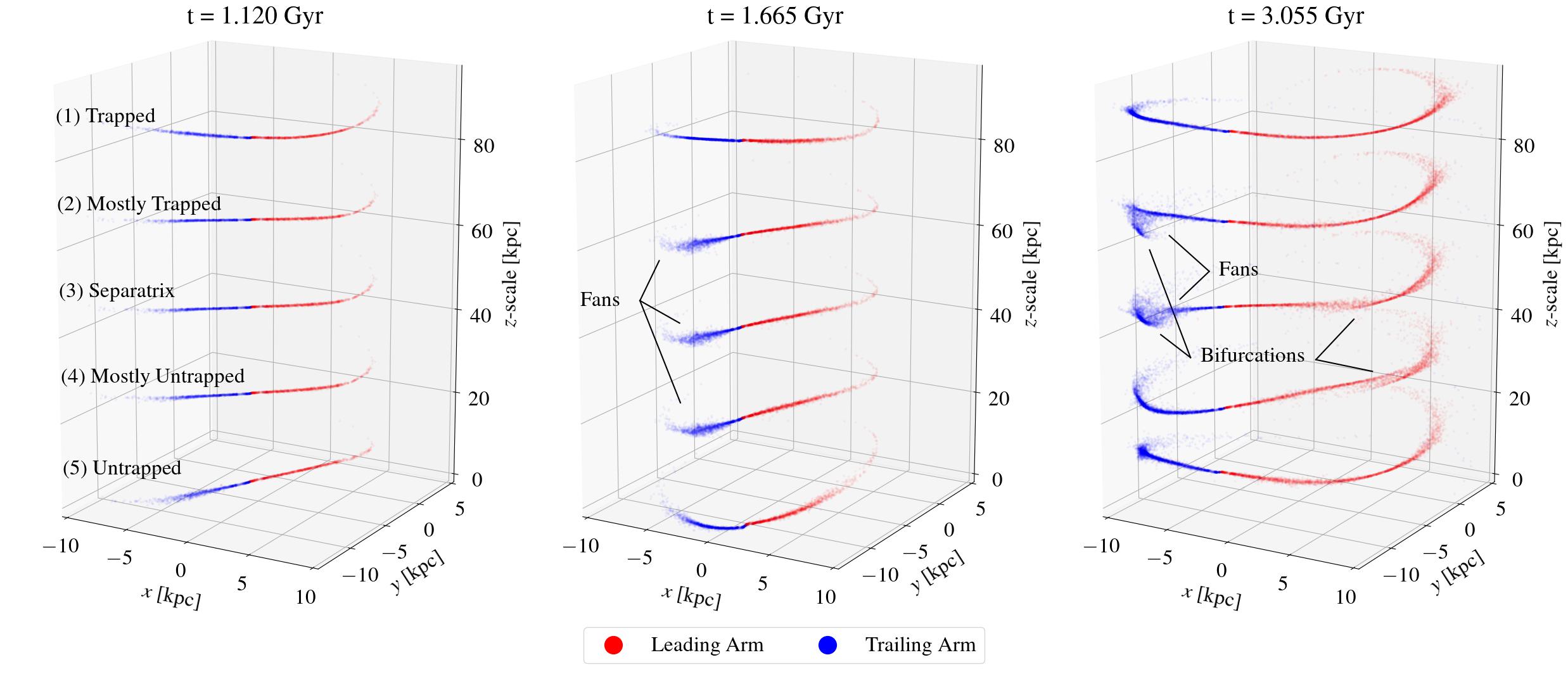}
    \caption{Three snapshots of the phase-space positions of the five $m=10^5$ M$_\odot$ streams introduced in \S\ref{subsec:4.1}, separated out from each other along the $z$-axis (the streams are evolved from the same $z$ position in the halo, and translated to higher/lower $z$ positions in these plots for visual clarity). The first panel demonstrates that initially, all the streams appear to be evolving as thin stellar streams. In the middle panel, roughly 0.5 Gyr later, the trailing arms (in blue) of the three streams that were initiated near the separatrix fan out compared to the fully trapped and the fully untrapped streams. The final panel shows that approximately 3 Gyr into their evolution, these three streams also develop bifurcations.}
    \label{fig:04}
\end{figure*}

\subsection{Simulation Setup}
\label{subsec:4.1}

We use the setup described in the previous section (i.e., the same potential and resonance shown in Figures \ref{fig:01} and \ref{fig:03}). From Figure \ref{fig:03}, it is tempting to choose the middle or bottom row, corresponding to an $m=10^7 - 10^9$ M$_\odot$ progenitor, given how the separatrix will cut through the stream's phase-space volume for a large subset of progenitor orbits. However, for these mass scales (especially at $m=10^9$ M$_\odot$), the phase-mixing time of the stream is considerably shorter than the libration time. As such, we focus in this section on detailed simulations of the $10^5$ M$_\odot$ stream.

Our main interest lies in understanding how the relative position of the stream with respect to the resonance affects the stream morphology. We model five individual streams, whose progenitors are all initialized along the line between the black `$\times$' and the bottom black triangle in the left-hand panel of Figure \ref{fig:01}:

\begin{enumerate}
  \item[(1)] A fully trapped stream, initialized from the black `$\times$' in Figure \ref{fig:01}.
  \item[(2)] A mostly trapped stream, initialized one $\delta v_\mathrm{sos}$ above the separatrix.
  \item[(3)] A stream initialized very close to the separatrix (just outside, so the progenitor is technically on an untrapped orbit, albeit one with a very long circulation period).
  \item[(4)] A mostly untrapped stream, initialized one $\delta v_\mathrm{sos}$ below the separatrix.
  \item[(5)] A fully untrapped stream, initialized from the black triangle below the resonance in Figure \ref{fig:01}.
\end{enumerate}

In all the cases discussed here, the streams are initialized near the apex of their $R-v_R$ contours in the SoS. As a result, the progenitors first pass by the \textit{unstable balance point} (i.e., where the two branches of the separatrix intersect) approximately $1/4$ $T_\ell$ into the simulation, at which point the effects of separatrix divergence will first become noticeable. In general, a stream on a near-resonant orbit should begin exhibiting the dynamical effects of separatrix divergence within $1/2$ $T_\ell$ of when the tidal stripping begins. The effects can also begin almost instantaneously -- if the tidal stripping happens to commence near the unstable balance point of the resonance.

\subsection{Fans, Bifurcations, and Asymmetries}
\label{subsec:4.2}

Figure \ref{fig:04} shows three snapshots of the five $10^5$ M$_\odot$ streams. In the first snapshot, taken before any of the progenitors reach the unstable balance point of the resonance, all five streams exhibit similarly thin morphologies. However, just 0.5 Gyr later, the trailing arms of the three streams nearest the separatrix fan out noticeably compared to both the fully trapped and the fully untrapped streams. Over the subsequent 1.5 Gyr, the streams that evolve near the separatrix become characterized by fans and bifurcations. The bifurcations are typically connected to the main stream track by a low-density fan of material, which may be difficult to detect observationally \citep[see, e.g.,][]{Bonaca2021OrbitalStreams}, making it challenging to associate between them and potentially leading to the incorrect conclusion that these are two separate streams.

The morphologies of the mostly trapped (2) and the mostly untrapped (4) streams appear at first very similar to each other, with both exhibiting fanned-out trailing arms in the second snapshot. However, by the 3 Gyr mark, the two streams take on clear differences, with the mostly trapped stream retaining its fanned out trailing arm and developing a bifurcation in the same arm, while the trailing arm of the mostly untrapped stream becomes thin again, and the \textit{leading} arm of that stream develops a bifurcation.\footnote{To be more precise, the more recently stripped particles that form the trailing arm of the mostly untrapped stream are no longer fanning out, while the particles that formed the original fanned out section of the trailing arm remain fanned out, but have evolved further away from the progenitor.}

Several questions arise from Figure \ref{fig:04}: does the timing of the onset of morphological differences match the expectation from theory? If separatrix divergence is the dynamical cause of the different morphologies, what explains the appearance of two distinct morphological effects (fans vs. bifurcations)? What are the dynamical reasons for the asymmetries between the leading and trailing arms in the second snapshot, and between the mostly trapped and mostly untrapped streams in the third snapshot?

\begin{figure*}
	\includegraphics[width=\textwidth]{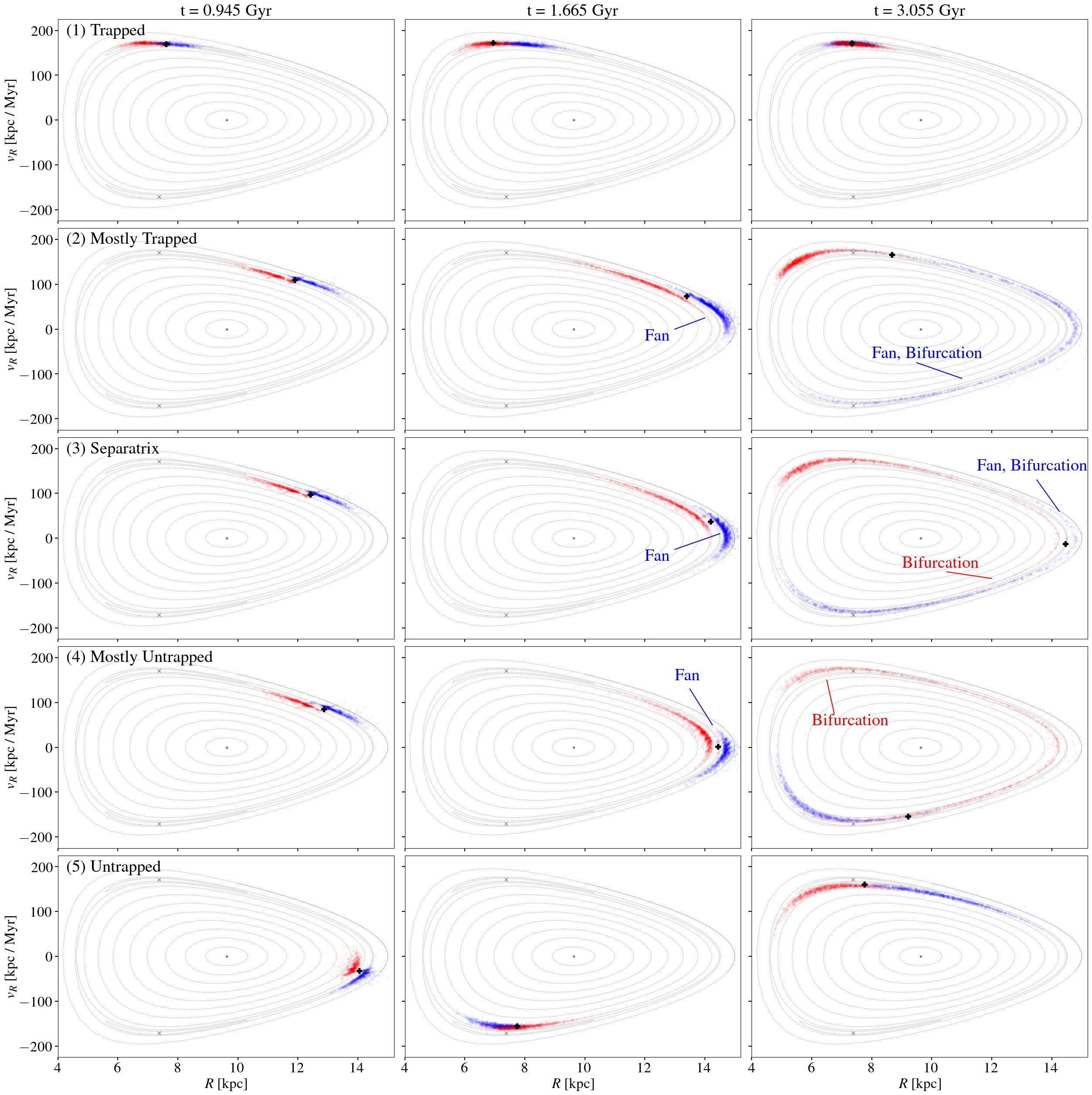}
	\caption{The five streams from Figure \ref{fig:04} plotted in SoS plots. The light gray contours and $\times$ markers are identical to those calculated for Figure \ref{fig:01}, and correspond to the energy and angular momentum of the progenitors of all five streams. Each column corresponds to one of the snapshots in Figure \ref{fig:04}, the bold black `+' reflects the progenitor's most recent consequent in the SoS before the timestamp of the snapshots in Figure \ref{fig:04}, and the red and blue points reflect the consequents of the orbits of the leading and trailing arm particles of each stream, just like in Figure \ref{fig:04}. The structure and evolution of each stream in the SoS plots can be related back to its morphology in Figure \ref{fig:04}, as shown in the labels in this figure; the color-coding of the labels corresponds to whether the feature belongs to the leading arm (red) or the trailing arm (blue).}
    \label{fig:05}
\end{figure*}

\subsection{Stream Evolution Visualized in SoS Plots}
\label{subsec:4.3}

To answer these questions, we visualize the \textit{entire stream} in SoS plots, shown in the grid in Figure \ref{fig:05}. It is important to reiterate that SoS plots are an imperfect tool for visualizing streams, because the stream particles differ from each other in their total energy and angular momentum. This issue is mitigated somewhat given the low progenitor mass in this case means the energy and angular momentum difference between stream particles is relatively small (as shown in the top row of Figure \ref{fig:03}). A second complication of using SoS plots to visualize an evolving object like a stream arises because individual particles cross the $z=0$ plane at different times, meaning the consequents of each orbit are not all recorded at the same time, and may not reflect the precise phase of the particle's orbit at a later time. Nevertheless, we find that visualizing the streams' evolution in SoS plots is particularly helpful for elucidating the underlying dynamics.

In order to create the SoS plots in Figure \ref{fig:05}, we plot each particle's \textit{most recent} consequent, thus providing an approximate sense of its orbital phase in the SoS. Given the orbital timescales, the time-spacing between consequents is approximately 200 Myr, or around $10\%$ of the shortest libration periods studied with these streams (as can be seen in the bottom-right panel of Figure \ref{fig:01}, $T_\ell \geq 2$ Gyr for all the progenitors studied in this section). The top row demonstrates how all the particles belonging to the fully trapped stream are confined to the top of the SoS, never straying anywhere close to $v_R = 0$ km/s. The bottom row provides the standard picture of stream evolution, in which all the untrapped particles evolve roughly along the invariant curve belonging to the progenitor, while slowly spreading out along that curve as the toy picture of standard phase-mixing would predict.

The first striking feature in the SoS plots of streams (2)-(4) is that the progenitor (shown in the bold black `+') is evolving more slowly here along the invariant curve than the untrapped case (the same is true for the comparison with the fully trapped case, except it is very hard to see the motion of the progenitor around the very short resonant invariant curve). Nonetheless, this slow evolution of the near-separatrix progenitors should not come as a surprise -- in fact, this is precisely the expectation based on the nature of $T_\ell$ near the separatrix shown in Figure \ref{fig:01}. Perhaps more puzzling is the fact that there doesn't appear to be a clear difference between the leading and the trailing arms in the central columns of panels corresponding to the 1.665 Gyr snapshot, whereas in phase-space the trailing arm is fanned out and considerably more diffuse than the leading arm; this, too, stems from the spread in libration times near the separatrix, as will be discussed further on.

\begin{figure*}
    \centering
    \includegraphics[width=0.82\textwidth]{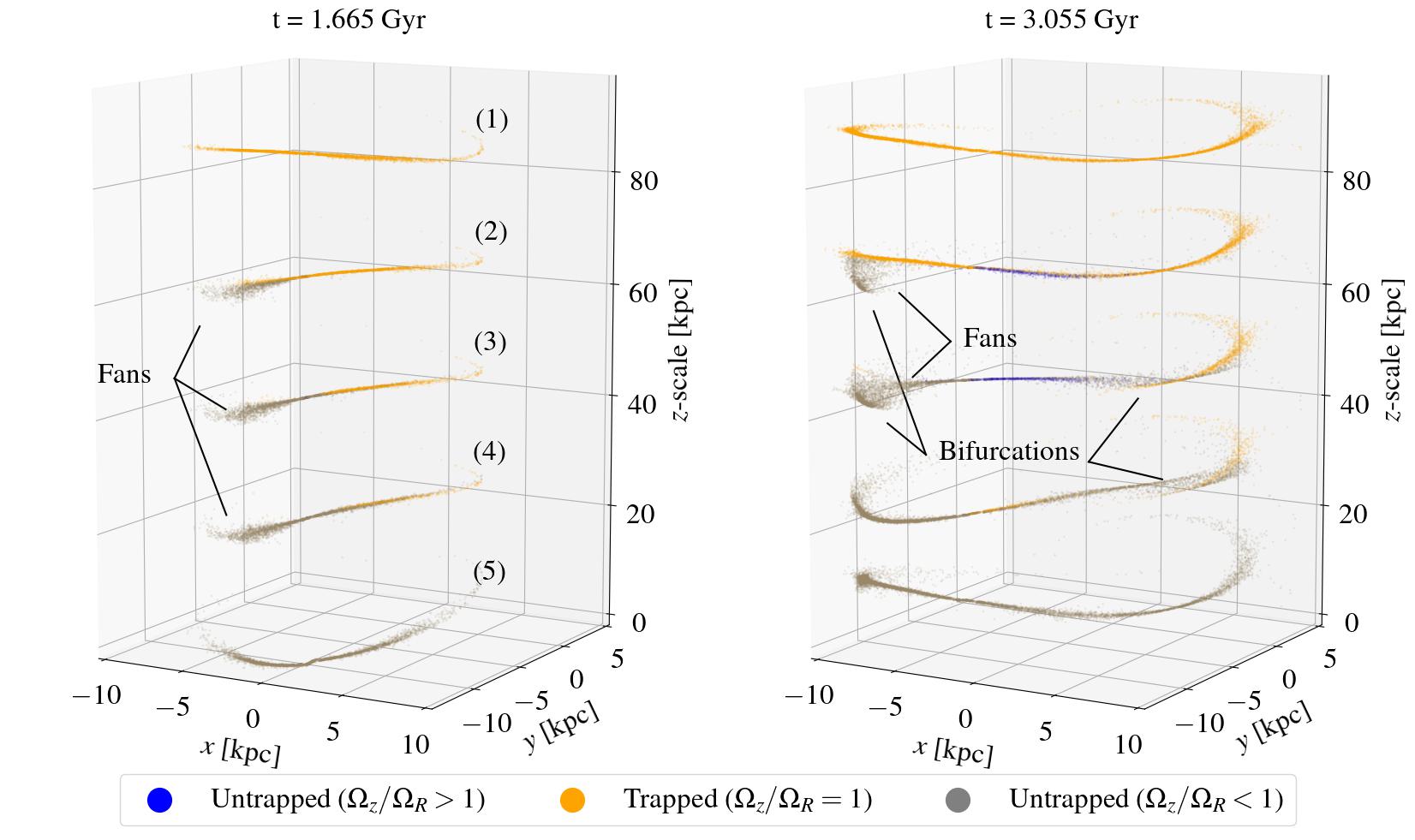}
    \includegraphics[width=0.82\textwidth]{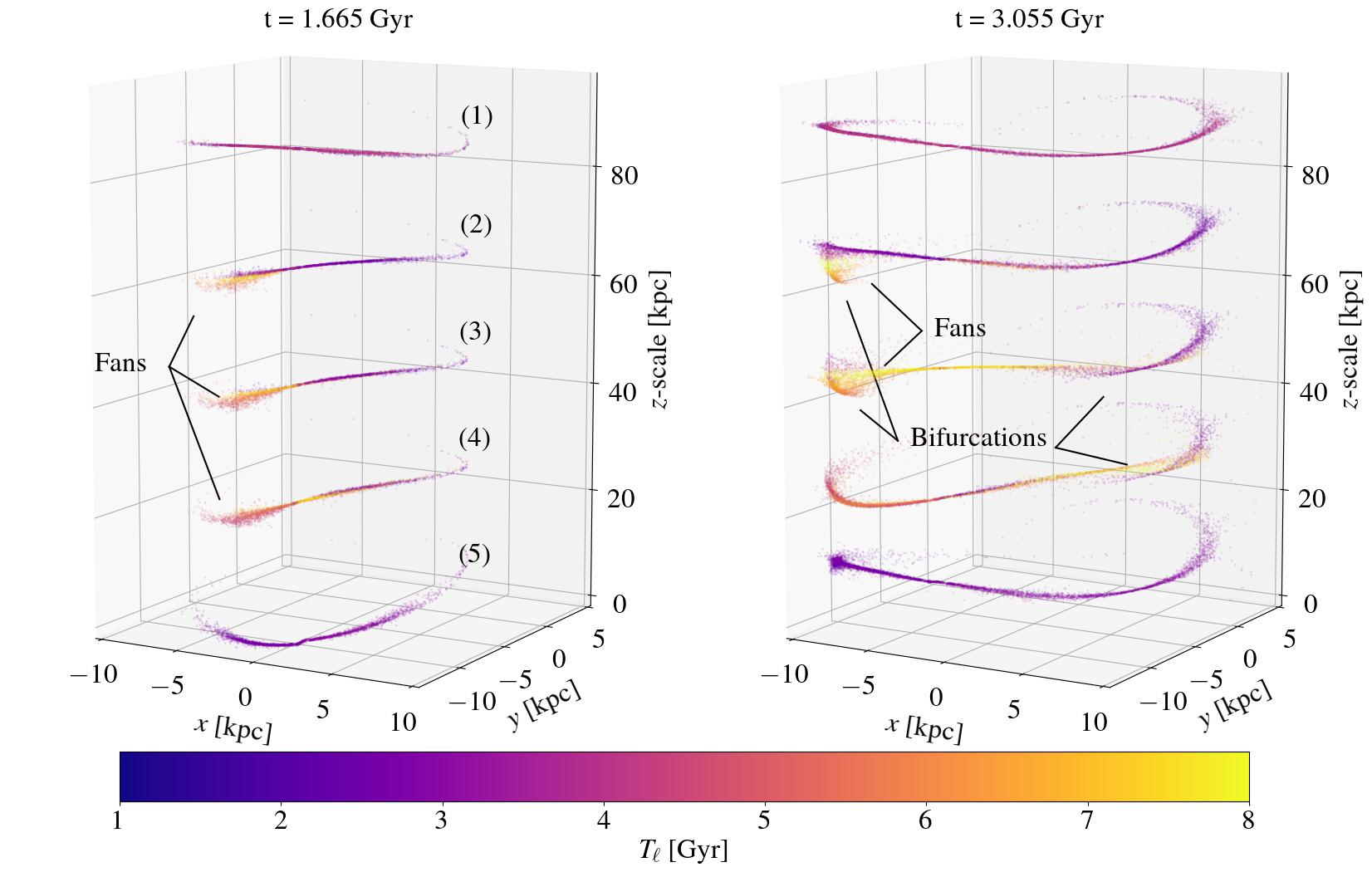}
    \caption{\textit{Top}: a recreation of the phase-space plots in the second and third panels of Figure \ref{fig:04}. Here, the color of each particle is based on the orbit family to which it belongs (as determined by the ratio of $\Omega_z$ and $\Omega_R$). The bifurcations appear to have a different color than the main stream track in the top-right panel (e.g., in stream (2), the bifurcation in the trailing arm is gray whereas the rest of that arm is orange), indicating that these components belong to different orbit families. At the same time, the fans highlighted in the top-left panel appear to mostly belong to the untrapped orbit family, suggesting that fans do not necessarily require particles from different orbit families in order to form. \textit{Bottom}: The same two plots, this time with the particle color determined by each particle's libration time $T_\ell$. The fans in the bottom-left panel are accentuated by a noticeable gradient in the libration time. Similarly, the fanned out debris structures between the main stream track and the bifurcations in the bottom-right panel are characterized by long libration times (i.e., they are yellow) compared to the bifurcations themselves. The physical source of the fanned out stream components is directly related to the large spread in these libration periods.}
    \label{fig:06}
\end{figure*}

Focusing on the third column of plots, corresponding to the 3.055 Gyr snapshot, the SoS plots provide a satisfying explanation for the bifurcations shown in the right-hand panel of Figure \ref{fig:04}. In the mostly trapped case, every particle in the leading arm (in red) appears to be on a trapped orbit, as evidenced by the fact that there are no red points in this SoS with $v_R < 0$. On the other hand, while some blue particles from the trailing blue arm also appear to be trapped (those that return to the top part of the SoS), others evolve past the unstable balance point to regions in the SoS where $v_R < 0$, indicating that they are not resonantly trapped. The separatrix thus divides the trailing arm into two groups of particles, belonging to two distinct orbit families, leading eventually to a bifurcation in the stream.

The reverse is true for the mostly untrapped stream: here, the final SoS plot indicates that the \textit{leading} (red) arm is split into two groups of trapped and untrapped particles, and indeed it is the leading arm that exhibits a bifurcation in Figure \ref{fig:04}. In the case of the separatrix stream, the separatrix appears to have separated between trapped and untrapped particles in \textit{both} the leading and the trailing arms, resulting in bifurcations in both arms.

\subsection{Relating Morphological Features to Properties of Near-Resonant Dynamics}
\label{subsec:4.4}

To better visualize the source of the bifurcation, we re-plot the second and third snapshots of in Figure \ref{fig:04} in top row of Figure \ref{fig:06}, where each particle's color is determined by its orbit family: resonantly trapped particles, with $\Omega_z / \Omega_R = 1$ are shown in orange, while untrapped particles are shown in either blue or gray, corresponding to whether the invariant curve of their orbit in the SoS plot is `above' the resonant orbit family ($\Omega_z / \Omega_R > 1$) or `below' it ($\Omega_z / \Omega_R < 1$). Using this coloration, the bifurcations shown in the panel on the top-right appear to correspond to different orbit families: stream (2) has an untrapped (gray) component in its trailing arm, stream (4) has a predominantly trapped component in its leading arm, and stream (3) exhibits multiple components corresponding to all three colors plotted in both its leading and trailing arms (we revisit these multiple components later in this section). This serves to further demonstrate that the bifurcations stem from a subset of particles belonging to a different orbit family.

On the other hand, the top-left panel (corresponding to $t=1.665$ Gyr) suggests that the fanned-out debris belongs predominantly to one orbit family (the fanned-out particles are mostly gray), meaning that unlike bifurcations, the mechanism for stream fanning does not necessarily require the particles to be divided by a separatrix into two orbit family groups.

The bottom row of Figure \ref{fig:06} shows the same streams, but this time with the particle color determined based on each particle's libration period $T_\ell$. Evidently, the dynamical source of the fanned out regions is a large spread in the libration periods, typical of the orbital region in the vicinity of the separatrix (as shown in the bottom right panel of Figure \ref{fig:01}). As the stream nears the unstable balance point associated with the resonance, the orbital evolution slows (said a different way, the location of the consequent on the SoS plot remains by the unstable balance point for a long time, similar to a pendulum reaching the unstable balance point at the top of its trajectory). Small differences in initial conditions can correspond to large variations in this slow libration period, meaning that each particle will move past the unstable balance point at a different rate. This leads to accelerated phase mixing, and results in the large spreads shown for both the leading and trailing arms in the final SoS plots for streams (2)-(4) in Figure \ref{fig:05}. This accelerated phase mixing is the primary reason for the stream fanning shown in Figure \ref{fig:06}.

The top right panel of Figure \ref{fig:06} highlights one final feature that warrants some attention: the most recently stripped particles in the leading arm of streams (2)-(4) (i.e., the right arm of the streams in this projection) are distinct from the other particles in that arm. In streams (2) and (3), some of the most recently stripped particles in the leading arm are on untrapped orbits (shown in blue), with $\Omega_z / \Omega_R > 1$, unlike the other particles in this arm, which are all trapped (in orange), or those in the trailing arm, some of which are trapped and some of which have $\Omega_z / \Omega_R < 1$ (gray). In the corresponding SoS plot in Figure \ref{fig:05} (third column, second row), the progenitor of this stream has begun moving back up on the top branch of the resonant invariant contour, and some of the tidally stripped particles are now being deposited on untrapped orbits \textit{above} the resonantly trapped island. A similar effect can be seen in stream (4), yet here, instead of the new particles belonging to the non-resonant orbit family with $\Omega_z / \Omega_R > 1$, they belong to the resonant orbit family at the bottom of the SoS (i.e., orbits that trace the upside-down reflection of the saucer orbit). In other words, 3 Gyr into its evolution, the leading arm of stream (4) has sourced two distinct resonant components (both shown in orange dots), which correspond to the two resonant islands associated with the 1:1 resonance shown in Figure \ref{fig:01}.

\begin{figure}
	\includegraphics[width=\columnwidth]{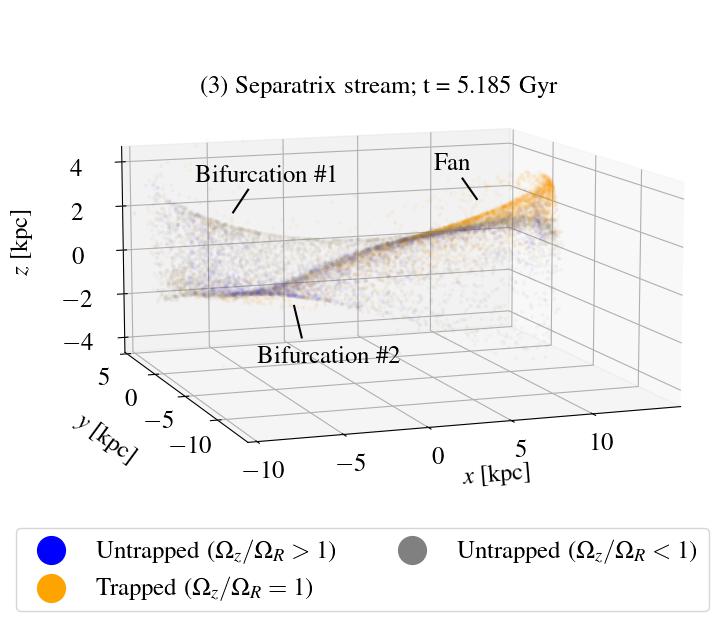}
	\caption{A snapshot of the leading arm of stream (3) from Figures \ref{fig:04}-\ref{fig:06} (a $10^5$ M$_\odot$ stream initiated very close to the separatrix). The stream was evolved for approximately five Gyr, and only particles belonging to the leading arm that were stripped in the first three Gyr of the stream's evolution are shown here. The morphology demonstrates how multiple components on distinct orbital tracks correspond to the different orbit families accessible to this stream.}
    \label{fig:07}
\end{figure}

In all of these cases, integrating the stream for long enough will lead to all of these components separating from each other and following distinct stream tracks. The separatrix stream (3) provides an excellent example of this in Figure \ref{fig:07}, in which two distinct bifurcations are formed from just the leading arm of the stream five Gyr into its evolution.

We conclude that separatrix divergence causes two distinct morphological effects for two different, albeit related, dynamical reasons. Bifurcations arise when a subset of stream particles is stripped into a different orbital family than the progenitor orbit, leading that group of particles to evolve along a different stream track (see groupings of blue/orange/gray particles in the top row of Figure \ref{fig:06}). Fans arise when the stream particles evolve close enough to a separatrix to exhibit a wide spread in their libration periods (see color gradient along the fanned-out debris in the bottom row of Figure \ref{fig:06}).

\section{Application to MW Streams}
\label{sec:obs}

In this section, we present a preliminary investigation of the feasibility of using separatrix divergence to probe the configuration of the MW halo's gravitational potential. While the previous section may provide compelling evidence that separatrix divergence can cause observable morphological features such as bifurcations and stream fanning, the setup was especially favorable in the sense that the streams were deliberately placed very close to the separatrix of a resonance with a relatively short libration time. Our aim in this section is to understand if separatrix divergence can reasonably affect the morphology of MW streams, given their observed phase-space positions, for a non-negligible subset of configurations of the MW's dark matter halo. Through this, we seek to demonstrate the plausibility of observing a stream that has undergone separatrix divergence in the MW, and therefore that this method has the potential to constrain the shape of the MW's halo.

We emphasize that this section does not purport to match or fit observed stream morphologies, nor do we aim to make any claims about constraining the shape of the potential to any specific values in this work. Our goal is simply to convince the reader that separatrix divergence may be an important effect for MW streams, and that it should be considered in future attempts to model the MW's potential through studying streams, especially when a stream's morphology exhibits features such as fans or bifurcations. We also hope to provide additional motivation for further observational searches that aim to either detect stream members that have strayed off of the predicted stream track, or to associate between physically separated stream components through their chemistry or ages.

For the purposes of the investigation in this section, we leverage the stream catalog from \citet{Malhan2022}. We filter for streams whose radial period is $\lesssim 0.5$ Gyr, in order to focus on streams that have completed enough orbits to raise their probability of experiencing separatrix divergence. For each of the selected streams (thirteen in total), we assume that the ``anchor point'' coordinates listed in \citet{Malhan2022} represent the present-day coordinates of the stream's progenitor.\footnote{This is, of course, not necessarily true, though it serves as a perfectly good assumption given our purpose is to demonstrate the general importance of separatrix divergence, not to precisely model and fit any of these streams} For the remainder of this section, we refer to the orbit defined by the ``anchor point'' as the progenitor orbit.

We seek to find configurations of the MW's dark matter halo potential that place the progenitor orbits near resonances with libration times shorter than the age of the Galaxy. We initialize an array of MW potentials, each consisting of a nucleus, a bulge, and a bar, as described in \S\ref{subsec:pot}, and a \citet{Lee} axisymmetric halo with varying axis ratio ($q = c/a$ in density) ranging from 0.6 to 1.4. We evaluate the fundamental frequencies of the progenitor orbits of the selected streams in each potential, and search for streams whose progenitor orbits lie near either the 1:1, 2:3, or 3:4 resonances (for the purposes of the demonstration in this section we limit our search to only these three low-order resonances, even though other resonances exist and may also potentially affect stream morphology). We find that the orbits of eight of the thirteen streams lie near one of these three resonances in a subset of the potentials: Ylgr, Kshir, M92, Palca, Gaia-9, Phoenix, Phlegethon, and GD-1. 

Next, we model each of these streams assuming a globular cluster-like progenitor\footnote{This generic value may deviate from the expected progenitor masses of several of these objects -- Palca, for example, is in fact likely to originate from a dwarf galaxy, not a globular cluster \citep{Li2022} -- but once again, our goal is to demonstrate the general feasibility of observing morphological manifestations of separatrix divergence, not to model and fit any of these streams precisely} with a mass of $m=4.4\times10^4$ M$_\odot$. We integrate the progenitor orbit of every stream 10 Gyr into the past in each potential, and using that phase-space position as a starting point, initialize a mock stream using the prescription in \S\ref{subsec:str}, generating a sample of 1,024 particles for each stream. We then evaluate the fundamental frequencies of each generated particle to determine whether or not it is resonantly trapped, and use this to calculate the spread of each stream's particles across the different orbit families. Our aim is to find a stream in which a sizeable fraction of the particles lie in a second orbit family for a certain range of potentials, suggesting that, in those potentials, it should undergo separatrix divergence. As a limiting criterion, we search for streams that include $>5\%$ of particles that belong to a different orbit family than that of the progenitor.\footnote{It is worth noting that this criterion could be broadened, given that stream fanning due to the spread in libration frequencies can also happen in the vicinity of the resonance even if none of the particles are trapped in the resonance, as discussed in the prior section.}

\begin{figure*}
    \includegraphics[width=\textwidth]{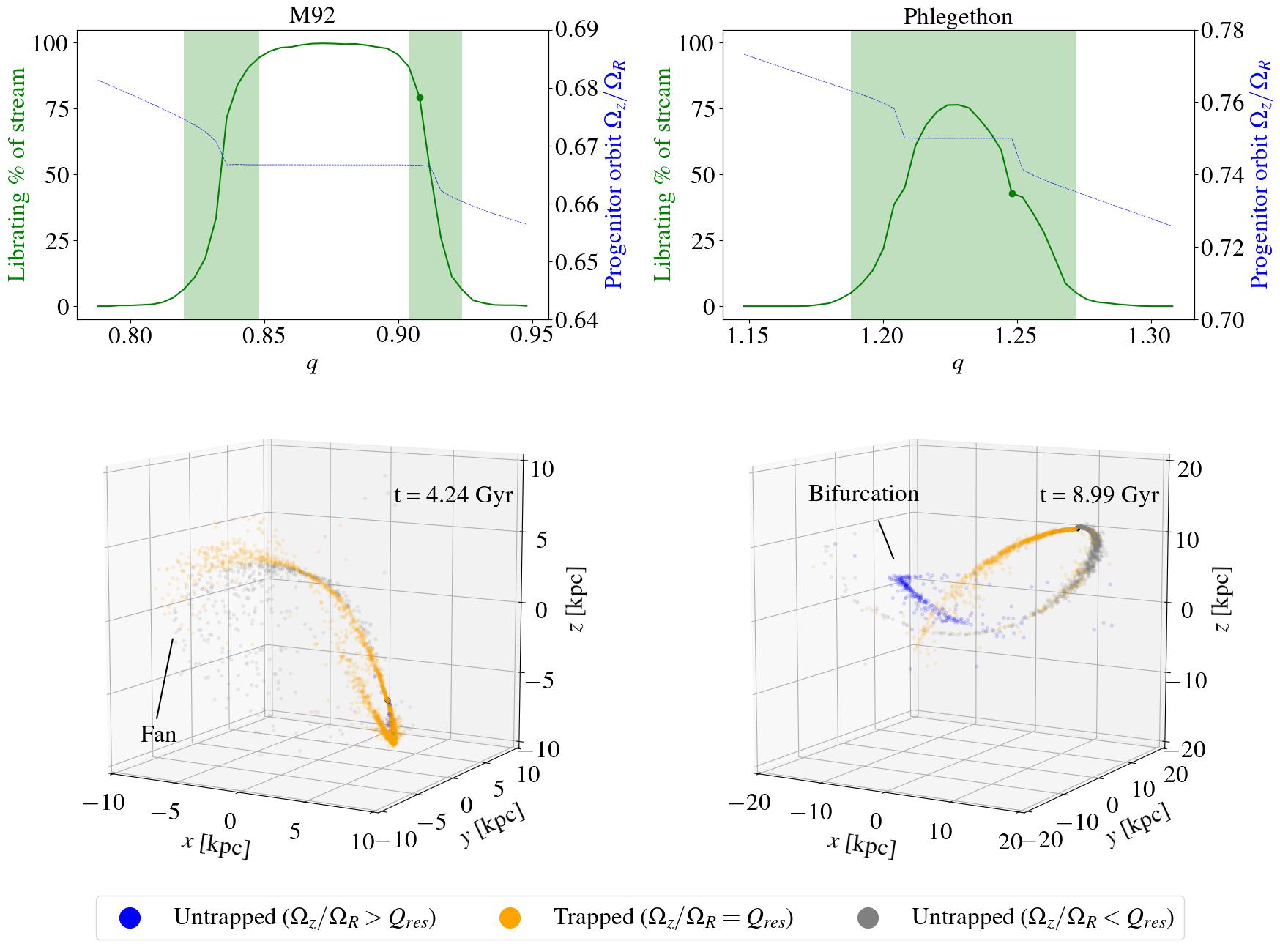}
    \caption{\textit{Top}: The solid green curve represents the percent of librating, or resonantly trapped particles in mock stream simulations of M92 (left panel) and Phlegethon (right panel), as a function of the density ratio parameter ($q=c/a$) of the halo potential. In M92, the stream encounters the $\Omega_z : \Omega_R = 2:3$ resonance around $q=0.875$. The highlighted green regions represent the potentials in which the stream particles are split across two orbit families (with at least 5\% of particles in a different orbit family than the progenitor), suggesting that in these potentials M92 could exhibit morphological deviations due to separatrix divergence. Phlegethon encounters a smaller resonantly trapped orbit family at the $\Omega_z : \Omega_R = 3:4$ resonance around $q=1.225$. Because in this case the stream occupies a greater volume in phase-space than the resonance, at no orbit is the stream fully trapped. As a result, the single highlighted green region extends across the entire range of potentials in which Phlegethon encounters the resonance. The blue line (and the associated secondary $y$-axis on the right of each panel) corresponds to the frequency ratio of the progenitor orbit as a function of $q$, demonstrating the range of potentials in which the progenitor encounters the resonance. Finally, the green dot marks the specific potential chosen for the visualizations in the bottom row of this figure. \textit{Bottom}: Mock stream simulations of M92 and Phlegethon in the two potentials marked by the green dots in the top row. Note that for M92, $Q_\mathrm{res} = 2/3$ while for Phlegethon $Q_\mathrm{res} = 3/4$. M92 exhibits a fan and Plegethon develops a distinct bifurcation within the lifespan of the Galaxy.}
    \label{fig:08}
\end{figure*}

We find that of the eight streams, M92 and Phlegethon provide the most promising cases for observing separatrix divergence within the range of potentials we initialized. As shown in the top row of Figure \ref{fig:08}, M92 is susceptible to separatrix divergence around $0.82 < q < 0.85$ and again at $0.9 < q < 0.92$, while a subset of Phlegethon is resonantly trapped when $1.19 < q < 1.27$.

As a test of these findings, we remodel these two streams in one selected potential within the ranges of interest described above -- this time with 6,000 particles -- and verify that the expected morphological effects of separatrix divergence begin occurring within 10 Gyr. As expected, both streams show clear evidence of separatrix divergence within this timeframe: the trailing arm of M92 becomes completely fanned out within 5 Gyr in the $q=0.908$ potential, while the leading arm of Phlegethon exhibits a clearly visible bifurcation by 9 Gyr in the $q=1.248$ potential, as shown in the bottom row of Figure \ref{fig:08} (the bifurcation in Phlegethon actually appears considerably earlier, around 5-6 Gyr, though here we show the snapshot at 9 Gyr in which the bifurcation is easily noticeable).

Between just these two streams, then, approximately 15\% of the potentials we investigated in this section could lead to one of the streams exhibiting morphological manifestations of separatrix divergence -- certainly a non-negligible portion of the total parameter space. Incorporating additional observed streams, as well as extending to higher order resonances, can likely push this percentage considerably higher.

A central observational challenge here lies in reliably identifying large numbers of sources that belong to a stream but have strayed off of the expected stream track. Given that most of the streams discussed here have fewer than 100 confirmed members each, current observational capabilities make it challenging to observe features like the ones discussed in this paper, though this will likely become more feasible with future data releases and the next generation of Earth- and space-based telescopes.

\section{Discussion}
\label{sec:disc}

The recent discoveries of many MW streams with features such as gaps, fans, and bifurcations, as well as orbital and chemical associations between seemingly separate streams, has led to several proposed mechanisms and explanations for inducing such features (a list to which, we argue, separatrix divergence should be added). In this section, we review some of these proposed mechanisms and discuss the similarities and differences between these mechanisms and separatrix divergence. In each of the following subsections, we seek to identify tests that will allow for differentiation between separatrix divergence and each of the other mechanisms.

\subsection{Standard Phase Mixing}

In the classical picture of stream evolution, standard phase mixing will eventually spread the stream stars across multiple wraps of the same orbit. If, while this happens, the stream maintains its coherence and still appears as a 1D filament of stars, the multiple wraps can appear to be a bifurcation from the main stream track, especially if the distribution along the stream track is non-uniform (and specifically if there is a large gap somewhere along the stream track that makes it harder to associate between the main portion of the stream and the apparent bifurcation).

This idea was first proposed to explain the bifurcation of the Sagittarius stream \citep{Fellhauer2006,Penarrubia2010,Ramos2021}. More recent studies have ruled this explanation out, in part because there is no clear reason for a long gap along the stream track between the bifurcation and the main portion of the stream, and in part because the orbital characteristics of the bifurcation appear to be qualitatively different from those of the rest of the stream \citep{Vasiliev2021}.

A similar configuration may also arise if the tidally stripped progenitor was a rotating disk galaxy, whose plane of rotation was misaligned with the Galactic disk \citep{Penarrubia2010}. More recent data regarding the precise configuration of the Sagittarius bifurcation remains consistent with this theory, however, the remnant core of the Sagittarius dwarf galaxy shows no evidence for internal rotation \citep{Penarrubia2011}. The reason for the bifurcation of Sagittarius thus remains a mystery.

In this regard, it is worthwhile noting that the mechanism described in this work does not appear to provide a natural explanation of the bifurcation of Sagittarius. In this case, the expected progenitor mass, and therefore the dynamical distribution of the stream's stars, are large enough to wash out any of the effects studied in this paper.

\subsection{Objects Falling in Together}

Another way to explain observations of two streams at different phases of the same orbit is to assume that rather than being part of the same object, \textit{two or more} objects fell into the MW together, and were deposited on (more or less) the same orbit at a different phase. Just like separatrix divergence, the orbital differences between the two objects can lead to the creation of two streams with very similar orbital characteristics, that are nonetheless far apart from each other on the sky. Several such associations between streams have been reported recently, including Jhelum and Indus \citep{Bonaca2019a} and Palca and AAU \citep{Li2021a}.

In many ways, the morphology brought about by this explanation is difficult to disentangle from separatrix divergence, given that both mechanisms lead to two distinct orbital clusterings. A promising approach for differentiating between these mechanisms involves studying the chemical properties of the two components, which can be accomplished thanks to recent work that has followed up chemically on stream detections \citep[see, e.g.,][]{Ji2020TheStreams,Martin2022TheEDR3,Malhan2022}. If the components are distinct, they likely originated from different objects that fell in together, such as a dwarf galaxy with an associated globular cluster. On the other hand, if they are similar chemically, separatrix divergence may be the more likely cause, because the separatrix is splitting a single, chemically uniform, cluster of stars into two groups with distinct orbital (but not chemical) distributions.

\subsection{Interactions with the Galactic Bar}

A rotating bar in the center of the Galaxy causes a periodic perturbation that can lead to resonant interactions with streams and cause the formation of gaps or stream fanning. Several examples of this mechanism include explanations for the truncation of Ophiuchus \citep{Price-Whelan2016b,Hattori2016ShepherdingStream} and the asymmetric fanning of Pal 5 \citep{Pearson2017, Bonaca2019b}.

From a dynamical perspective, the resonant effects described in this paper and periodic perturbations from a barred distribution are very similar to each other, since they both pertain to resonant interactions. Here, we have described trapping resonances in static axisymmetric potentials that result from a coupling between motion along two different axes when the orbital frequencies along these axes are commensurable. For resonances with the bar, the resonance is instead between one of the orbital frequencies of the stream and the Galactic bar's pattern speed (or frequency).

Resonances with the bar lead to significantly more pronounced effects when a stream evolves on a prograde orbit \citep{Pearson2017}. If the stream's orbit is retrograde, the bar has little effect on the stream's morphology, so separatrix divergence may be a more likely explanation for any observed features in retrograde streams.

\subsection{Subhalo Encounters}

One of the more exciting directions of study related to stellar streams involves the attempt to discover evidence for Galactic substructure, and specifically CDM subhalos \citep{Johnston2002, Ibata2002, Yoon2011, Carlberg2012, Erkal2016a, Bovy2017}. When a dense object passes close to a stream, it can perturb the orbits of a subset of the stream particles and lead to the formation of a gap and a bifurcation/spur. Studying the exact nature of the gap can allow for the characterization of the object that caused it, including its mass, concentration, and orbital path with respect to the stream.

The most prominent example of this mechanism in the recent literature is the explanation of the gap and spur in GD-1 as the outcome of a dark matter subhalo fly-by \citep{Price-Whelan2018b, Bonaca2019}. Subhalo encounters have also been posited as possible causes of underdensities in Pal 5 \citep{Erkal2017} and Phoenix \citep{Tavangar2022FromStream}.

We offer two qualitative tests to differentiate between subhalo interactions and separatrix divergence:
\begin{enumerate}
    \item Gap+bifurcation vs. bifurcation: a localized encounter with another body such as a subhalo will typically lead to the formation of a gap and an accompanying bifurcation -- the stars that were removed from the gap are what forms the bifurcation -- similar to the observed morphology of the GD-1 stream \citep{Price-Whelan2018a}. On the other hand, separatrix divergence acts in a more uniform manner on the stream throughout its evolution. A bifurcation from separatrix divergence leads to a separate component of the stream that comprises stars drawn more or less uniformly from the main stream track. As a result, there is no reason for a bifurcation from separatrix divergence to have a gap associated with it, like one might expect for a subhalo interaction.
    \item A distinct orbital path: bifurcations caused by separatrix divergence must, by definition, trace a distinct orbital path in the Galactic potential (typically, one component will be on a resonantly trapped orbit and the other will be on an untrapped orbit, as discussed in \S\ref{subsec:4.4}). These bifurcations should be long-lived and should maintain their coherence long after they form. Unlike this, the bifurcations caused by subhalo encounters need not have any orbital coherence. In other words, the stars removed from the gap are all deposited on different orbits from each other, and may only appear as short-lived features like spurs or loops of stars, before rapidly phase mixing away. This test serves as a relatively straightforward method that can be carried out by integrating the observed stream particles in a reasonable MW potential, as long as there is sufficient kinematic information for both the main stream and the bifurcation \citep[e.g.,][]{Bonaca2020High-resolutionSagittarius}.
\end{enumerate}

Both of these tests rely on the fact that subhalo encounters are expected to be relatively local in nature, affecting only a subset of the stream. Larger encounters, such as the possible interaction between AAU and Sagittarius proposed in \citet{Li2021a}, will have a more global influence on the stream and may lead to large subsets of particles being deposited on similar orbits such that they don't necessarily lose coherence and phase mix away. In some cases, the results of encounters with large perturbers can also take on the appearance of the bifurcations discussed here \citep[see, e.g.,][]{Li2021a,Dillamore2022TheStreams}.

Finally, repeated interactions with smaller scale dark matter substructure can also cause streams to become more diffuse. In CDM, repeated encounters with low-mass subhalos are not expected to greatly affect stream morphology \citep{Yoon2011}, but in other models of dark matter, such as Fuzzy Dark Matter, these kinds of interactions can greatly impact stream morphology \citep{Hui2017,Dalal2020}. One may therefore wonder whether the stream fanning discussed above may be confused with such an effect. However, given the global and isotropic nature of the encounters with dark matter substructure, the entire population of MW's stellar streams would be equally affected by them, rather than just those that happen to be orbiting in a certain region of phase-space near a resonance. Furthermore, there would be no inherent reason for the effect to preferentially occur in only one of the two arms of a stream, as we showed to be the case with separatrix divergence.

\section{Summary and Conclusion}
\label{sec:conc}

We have studied the evolution of simulated stellar streams on near-resonant orbits and demonstrated that their morphologies exhibit fans or bifurcations if their orbits are close enough to a separatrix. Using a combination of orbital frequency analysis and surface of section plots, we have developed an efficient numerical approach for measuring the libration frequencies of near-resonant orbits. We use this technique to relate a stream's size and orbital distribution to the width of the resonance, and to study the dynamical processes that lead to the formation of fan and bifurcations.

We show that streams tend to bifurcate when a subset of stars is tidally stripped into a different orbit family compared to that of the progenitor's orbit, whereas they fan out as a result of the steep gradient in libration times of orbits near the separatrix (particularly in the untrapped orbital regions close to the separatrix). Lastly, we demonstrate the plausibility of separatrix divergence having these effects on actual MW streams, by showing how, in certain axisymmetric choices for the dark matter halo, separatrix divergence can cause observable morphological features in both the M92 and Phlegethon stellar streams, over the course of just a few Gyr.

A key open question relates to the highly delicate nature of the secular resonances studied in this work. Studying the behavior of streams on near-resonant orbits in the presence of effects that may disrupt the delicate balance that leads to secular dynamical effects would therefore serve as an important supplement to these results. Such effects include global evolution of the Galactic potential \citep[see, e.g.,][]{Vasiliev2021, Garavito-Camargo2021QuantifyingExpansions}, which may either displace a stream or cause resonances to shift around in phase-space, and the influence of diffusive processes, that add a degree of stochasticity to the otherwise regular orbits in the potentials studied here and can thus cause orbits to move into and out of resonantly trapped regions \citep{Hamilton2022GalacticFriction}.

These complications notwithstanding, we believe separatrix divergence merits consideration alongside other mechanisms for disrupting streams. As the catalog of observed streams in the MW continues to grow, a more robust understanding of these effects, how they interact with each other, and how they may be distinguished from one another, will become increasingly important. Furthermore, the day is nearing when next generation telescopes will begin detecting large catalogs of streams in other galaxies \citep{Pearson2019, Pearson2022}. An effect like separatrix divergence, which can largely be identified based on the morphology of streams without needing additional data, may therefore be particularly valuable for studying the shapes of external galaxies.

The methods presented in this paper and in \citetalias{Yavetz2021} serve as the basis for producing maps of galactic potentials and identifying regions of phase-space volume in which stellar streams undergo unique morphological evolution. Studying the extent of these regions and quantifying the fraction of the total phase-space volume they occupy is a key question that we hope to address in future work.

\begin{acknowledgments}

We wish to thank Martin Weinberg, Robyn Sanderson, Ting Li, Adam Wheeler, Monica Valluri, David Hogg, Melissa Ness, and Greg Bryan for helpful comments, suggestions, and advice. TDY wishes to further thank everyone in the \textit{Milky Way Stars} Group at Columbia University and the CCA Dynamics Group, for reviewing and critiquing various elements of this work in the year leading up to its publication.

TDY is supported through the NSF Graduate Research Fellowship (DGE-1644869) and the Corning Glass Works Foundation Fellowship at the Institute for Advanced Study. KVJ is partially supported by the National Science Foundation under grant no. AST-1715582. SP is supported by NASA through the NASA Hubble Fellowship grant \#HST-HF2-51466.001-A awarded by the Space Telescope Science Institute, which is operated by the Association of Universities for Research in Astronomy, Incorporated, under NASA contract NAS5-26555.
This work was supported by a grant from the Simons Foundation (816048, CH).

We thank the Center for Computational Astrophysics at the Flatiron Institute for support and space to conduct this work. The Flatiron Institute is supported by the Simons Foundation.

\end{acknowledgments}

\software{Astropy \citep{Robitaille2013,Price-Whelan2018b},  
          Gala \citep{Price-whelan2017}, 
          Matplotlib \citep{Hunter2007}, 
          Numpy \citep{scikit-learn},
          Scipy \citep{Virtanen2019},
          Superfreq \citep{Price-Whelan2015}
          }

\newpage{}

\appendix

\section{Representation of a Triaxial Halo Potential}
\label{sec:app-a}

In order to model non-spherical halos, we rely on the first-order approximation for  a triaxial NFW potential from \citet{Lee}, as implemented in \texttt{gala} \citep{Price-whelan2017}. This formulation is advantageous in that the axis ratios are all specified in terms of the density, thus ensuring that the density is never negative (as can sometimes be the case when the triaxiality is specified in terms of the axis ratios of the potential itself). The potential is expressed using spherical coordinates, with the dimensionless variable $u$ representing the ratio between the radius and the scale radius ($u = r / r_s$):
\begin{equation}
    \label{eq:pot}
    \Phi(u, \phi, \theta) = \frac{v_c^2}{A}\bigg[F_1(u) + \frac{e_b^2 + e_c^2}{2}F_2(u) + \frac{(e_b\sin\theta\sin\phi)^2 + (e_c\cos\theta)^2}{2} F_3(u)\bigg] \ ,
\end{equation}
with $v_c$ representing the circular velocity at the scale radius, and,

\begin{eqnarray}
    A & = & \bigg(\ln2 - \frac{1}{2}\bigg) + \bigg(\ln2 - \frac{3}{4}\bigg)(e_b^2 + e_c^2) \ , \\
    F_1(u) & = & -\frac{1}{u}\ln(1 + u) \ , \\
    F_2(u) & = & -\frac{1}{3} + \frac{2u^2 - 3u + 6}{6u^2} + \bigg(\frac{1}{u} - \frac{1}{u^3}\bigg)\ln(1+u) \ , \\
    F_3(u) & = & \frac{u^2 - 3u - 6}{2u^2(1+u)} + \frac{3}{u^3}\ln(1+u) \ , \\
    e_b & = & \sqrt{1 - (b/a)^2} \ , \\
    e_c & = & \sqrt{1 - (c/a)^2} \ .
\end{eqnarray}

Throughout the paper, we set $v_c = 170$ km / s, $r_s = 15.62$ kpc, and $b/a = 1$. We use an oblate halo potential with $c/a = 0.8$ in Sections \ref{sec:res1} and \ref{sec:res2}, whereas in Section \ref{sec:obs} we vary the degree of oblateness. The other three components (nucleus, bulge, and disk) are always the same, and follow the parameters specified in \texttt{MilkyWayPotential} in \texttt{gala} \citep{Price-whelan2017, Bovy2014a}.

\

\section{Secular Perturbation Theory and the Scale-Free Nature of Separatrix Divergence}
\label{sec:app-b}

In the framework of secular perturbation theory, the Hamiltonian describing
motion of a test particle in the vicinity of a resonance can be reduced to that of a pendulum:
\begin{equation}
    \label{eq:app_ham}
    H(\theta_s, I) = \frac{1}{2}GI^2 - F\cos (k\theta_s).
\end{equation}
Here, the canonical phase-space coordinates are $I \equiv J_s - J_{s0}$, which is the deviation of the test particle's \textit{slow action} $J_s$ from its resonant value $J_{s0}$, and the corresponding \textit{slow angle} $\theta_s$, which does not change with time for a perfectly resonant particle.  
We have also assumed that a single Fourier component $k$
dominates the expansion of the perturbing Hamiltonian; in the present case we may set $k=2$ \citep{Binney1993}.
The other quantities are constant coefficients evaluated on resonance: $G\equiv \partial^2H_0 / \partial J_s^2\vert_{J_{s0}}$ measures the curvature of the unperturbed Hamiltonian, and $F\equiv -2 \vert H_k(J_{s0}) \vert $ measures the strength of the perturbation.  (It is important to note that we have implicitly fixed the \textit{fast actions} $\mathbf{J}_f$ when writing down this pendulum Hamiltonian; changing $\mathbf{J}_f$ would change $J_{s0}$, $G$ and $F$).
The pendulum's phase space structure is then completely determined by two key quantities: 
the libration frequency for small oscillations around the resonance $\omega_\ell \equiv \sqrt{kFG}$, and the separatrix half-width $I_\mathrm{h} \equiv 2\sqrt{F/G}$.
See \citetalias{Yavetz2021} and \cite{Hamilton2022GalacticFriction} for a more complete treatment of this formalism.\footnote{We note here a typo in \S2.3 of \citetalias{Yavetz2021}; the separatrix corresponds to $\Delta \hat{H} = F$, not to $\Delta \hat{H} = 0$, which means that the separatrix half-width is given by $I_\mathrm{h} = 2\sqrt{F/G}$, not $\delta J_{s\mathrm{max}} = \sqrt{2F/G}$. Also, the libration frequency for small oscillations around a resonance is $\omega_\ell = \sqrt{kFG}$ (rather than $\sqrt{FG}$).
These modifications do not change any of the basic conclusions, though.}

For an ensemble of test particles with fixed $\mathbf{J}_f$, the evolution under the Hamiltonian $H(\theta_s, I)$ is governed by 
\begin{equation}
    \label{eq:app_pm1}
   \frac{\mathrm{d}f}{\mathrm{d}t}  =  \frac{\partial f}{\partial t} + \{f,H\} = 0,
\end{equation}
where $f(\theta_s, I, t)$ is the distribution function of the ensemble, and $\{ f, H \} \equiv (\partial_{\theta_s} f) \, (\partial_I H) - (\partial_I f) \, (\partial_{\theta_s} H) $ is a Poisson bracket.
Substituting (\ref{eq:app_ham}) into (\ref{eq:app_pm1}), we obtain:
\begin{equation}
    \label{eq:app_pm2}
    \frac{\partial f}{\partial t} + GI\frac{\partial f}{\partial \theta_s} - kF\sin(k\theta_s)\frac{\partial f}{\partial I} = 0.
\end{equation}
Let us now introduce the dimensionless time
$\tau \equiv \sqrt{kFG}\, t = \omega_\ell\, t$
and the dimensionless slow action 
$j \equiv I \sqrt{G/(kF)} = (2/\sqrt{k})I/I_\mathrm{h}$.
Then we can rewrite (\ref{eq:app_pm2}) as:
\begin{equation}
    \label{eq:dfdt_dimensionless}
    \frac{\partial f}{\partial \tau} + j\frac{\partial f}{\partial \theta_s} - \sin (k\theta_s)\frac{\partial f}{\partial j} = 0.
\end{equation}
In this equation, all quantities are dimensionless (up to the normalization of $f$, which is arbitrary since this is a linear equation). As a consequence, results gleaned from analysis of Equation \ref{eq:dfdt_dimensionless} are
applicable to any dynamical system in which resonant trapping plays a role, and do not just apply to the specific galactic scales chosen for this paper.

\bibliography{references}{}
\bibliographystyle{aasjournal}

\end{document}